\documentclass[format=acmlarge, review=false, screen=true]{acmart}

\usepackage{amssymb}
\usepackage{amsmath}
\usepackage{graphicx}
\usepackage{subcaption}
\usepackage{multicol}
\usepackage{algorithm}
\usepackage{algorithmic}
\usepackage{listings}
\usepackage{color}
\usepackage[english]{babel}
\usepackage{url}
\usepackage{multirow}
\newcommand{\hide}[1]{}

\setcopyright{rightsretained}

\begin{document}
\title{False Information on Web and Social Media: A Survey}

\author{Srijan Kumar}
\affiliation{
\institution{Computer Science, Stanford University, USA}
}
\email{srijan@cs.stanford.edu}
\author{Neil Shah}
\authornote{Dr. Shah is now at Snap Inc.}
\affiliation{
\institution{Computer Science, Carnegie Mellon University, USA}
}
\email{neil@cs.cmu.edu}

\begin{abstract}
False information can be created and spread easily through the web and social media platforms, resulting in widespread real-world impact.
Characterizing how false information proliferates on social platforms and why it succeeds in deceiving readers are critical to develop efficient detection algorithms and tools for early detection.
A recent surge of research in this area has aimed to address the key issues using methods based on feature engineering, graph mining, and information modeling.
Majority of the research has primarily focused on two broad categories of false information: opinion-based (e.g., fake reviews), and fact-based (e.g., false news and hoaxes).
Therefore, in this work, we present a comprehensive survey spanning diverse aspects of false information, namely (i) the actors involved in spreading false information, (ii) rationale behind successfully deceiving readers, (iii) quantifying the impact of false information, (iv) measuring its characteristics across different dimensions, and finally, (iv) algorithms developed to detect false information.
In doing so, we create a unified framework to describe these recent methods and highlight a number of important directions for future research.\footnote{A previous version of this survey will appear in the book titled Social Media Analytics: Advances and Applications, by CRC press, in 2018.}
\end{abstract}

\keywords{misinformation, fake news, fake reviews, rumors, hoaxes, web, internet, social media, social networks, bad actors, bots, propaganda, conspiracy, knowledge bases, e-commerce, disinformation, impact, mechanism, rationale, detection, prediction}

\maketitle

\section{\textbf{Introduction}}

The web provides a highly interconnected world-wide platform for everyone to spread information to millions of people in a matter of few minutes, at little to no cost~\cite{berners2010world}.
While it has led to ground-breaking phenomenon such as real-time citizen journalism~\cite{hermida2010twittering}, at the same time it has led to increased visibility and impact of both true and false information~\cite{mendoza2010twitter}.
False information on the web and social media has affected stock markets~\cite{bollen2011twitter}, slowed responses during disasters~\cite{gupta2013faking}, and terrorist attacks~\cite{fisher2016pizzagate,starbird2014rumors}.
Recent surveys have alarmingly shown that people increasingly get their news from social media than from traditional news sources~\cite{perrin2015social,shearer2017news}, making it of paramount importance to curtail false information on such platforms.
With primary motives of influencing opinions and earning money~\cite{nprfake,makhija2015flipkart,luca2016fake,smith2016macedonia}, the wide impact of false information makes it one of the modern dangers to society, according to the World Economic Forum~\cite{howell2013digital}.
Understanding the reasons for why and how false information is created is important to proactively detect it and mitigate its impact.
In this survey, we review the state of the art scientific literature on false information on the web and social media to give a comprehensive description of its \textbf{\textit{mechanisms, rationale, impact, characteristics}}, and \textbf{\textit{detection}}.
While recent surveys have focused on fake news in social media~\cite{shu2017fake,zubiaga2017detection}, the current survey broadly focuses on three types of false information on the web and social media---\textbf{fake reviews in e-commerce platforms, hoaxes on collaborative platforms, and fake news in social media.}

For ease of explanation and understanding, we categorize false information based on its \emph{intent} and \emph{knowledge} content.
We also broadly focus on false information that is public and targets many people at the same time, such as false reviews or fake tweets, as opposed to targeted false information as in cases of scam.
According to intent, false information can be categorized as \emph{misinformation}, which is created \textit{without} the intent to mislead, and \emph{disinformation}, which is created \textit{with} the intent of misleading and deceiving the reader~\cite{fallis2014functional, hernon1995disinformation}.
Both have negative influences, but the latter is arguably more dangerous as its creator's primary aim is expressly malicious.
Based on knowledge, false information is categorized as \textbf{\emph{opinion-based}}, where a unique ground truth does not exist as in cases of reviewing products on e-commerce websites, or as \textbf{\emph{fact-based}}, which consists of lies about entities that have unique ground truth value~\cite{thomas1986statements}.
We study both these types of false information in this survey.

The \textbf{\textit{bad actors involved}} in creating and spreading false information on a large scale use armies of fake accounts, such as bots and sockpuppets~\cite{ferrara2016rise,kumar2017army,shah2017manyfaces,subrahmanian2016darpa}.
These accounts are synthetically created or compromised~\cite{shah2017manyfaces}, controlled by a single underlying entity, and engineer the spread of false information through social networks~\cite{bakshy2011everyone,cheng2014can}, with the aim of creating an illusion of public consensus towards the false pieces of information.
Bots operate on a large scale for two purposes: first, to spread the same content, e.g., by retweeting, to a large audience, and second, by following each other to increase the social status of the accounts and apparent trustworthiness of information~\cite{ferrara2016rise,subrahmanian2016darpa}.
Sockpuppet accounts engage with ordinary users in online discussions and agree with each other to amplify their point of view and oppose those who disagree with the information~\cite{kumar2017army}.
Alarmingly, bots and sockpuppets hold central locations in information networks and therefore, are in key positions to spread false information.
We dig deep into the mechanisms of false information and the actors involved in Section~\ref{sec:mechanisms}.

False information would be ineffective if readers were able to easily identify that it is false and just discard it.
However, the \textbf{\textit{rationale behind successful deception}} by false information is evident from several research studies which show that humans are actually poor judges of false information~\cite{kumar2016disinformation,ott2011finding,perez2017automatic,yao2017automated}.
Specifically, humans are able to identify false information with accuracies between 53\% and 78\% across experiments with different types of false behaviors, including hoaxes, fake reviews, and fake news.
Both trained and casual readers get fooled into believing false information when it is well written, long, and is well-referenced.
Moreover, technological effects such as content personalization can lead to the creation of ideological echo chambers, so that people would receive the same false information multiple times through different channels and could even make it ``go viral''.
Biases in information consumers (e.g., confirmation bias), lack of education, and low media consumption lead to people being deceived by false information.
Further details of the rationale of deception using false information are explained in Section~\ref{sec:rationale}.

The spread of false information can have \textbf{\textit{far-reaching impact}}.
Several research studies have measured the impact of false information in social media in terms of user engagement metrics, such as the number of likes, reshares, and pre-removal lifetime, for hoaxes~\cite{kumar2016disinformation}, fake news~\cite{gupta2013faking,silverman2016analysis}, and rumors~\cite{friggeri2014rumor}.
They found that a small fraction of false information stories is highly impactful---they are liked, shared, and commented on more, generate deeper cascades of reshares than true information pieces, survive for a long time, and spread across the web.
This high engagement of false information with readers shows the degree of impact it can have on public opinion and ideological perception.
We discuss the impact of false information in detail in Section~\ref{sec:impact}.

In response, considerable research has been conducted to both investigate and use the \textbf{\textit{characteristics of false information}} to predict the veracity of new content.
Fake reviews~\cite{jindal2008opinion,li2017bimodal,mukherjee2012spotting,ott2011finding,sandulescu2015detecting,shah2016edgecentric}, hoaxes~\cite{kumar2016disinformation}, and fake news~\cite{bessi2016social,friggeri2014rumor,horne2017fake,perez2017automatic,silverman2016analysis,zubiaga2016analysing} have been characterized using their textual content, temporal features, ratings, references, user properties, network properties, spreading behavior, and mitigation behavior.
Along these characteristics, false information differs significantly from real information.
For instance, the text is generally longer, more exaggerated, and more opinionated compared to real reviews.
Temporally, fake reviews are created in short bursts, i.e., several fake reviews are usually written by the same account or group of accounts in a short time period.
The users who write these fake reviews and hoaxes are typically relatively new accounts with fewer reviews, and their local networks are often highly dense or overlapping.
Additionally, majority of fake news is spread by a very small number of users and it spreads rapidly during its initial release, before it is even debunked.
The characteristics of false information are discussed extensively in Section~\ref{sec:characteristics}.

\begin{table}
\begin{center}
\begin{tabular}{|c|c|}
\hline
\textbf{Social Platform} & \textbf{Research papers} \\\hline
\multirow{5}{*}{Twitter} & 
Bessi et al.~\cite{bessi2016social}, Ferrara et al.~\cite{ferrara2016rise}, Gupta et al.~\cite{gupta2013faking}, Howard et al.~\cite{howard2016bots}, \\
& Jin et al.~\cite{jin2016news,jin2013epidemiological}, Kim et al.~\cite{kim2018leveraging}, Mendoza et al.~\cite{mendoza2010twitter}, Mitra et al.~\cite{mitra2017parsimonious,mitra2015credbank}, \\
& Nied et al.~\cite{nied2017alternative}, Qazvinian et al.~\cite{qazvinian2011rumor}, Ruchansky et al.~\cite{ruchansky2017csi}, Shah et al.~\cite{shah2017manyfaces}, \\
& Shao et al.~\cite{shao2016hoaxy,shao2017spread}, Starbird et al.~\cite{starbird2014rumors,starbird2017examining}, Subrahmanian et al.~\cite{subrahmanian2016darpa}, Tripathy et al.~\cite{tripathy2010study}, \\
& Vosoughi et al.~\cite{vosoughi2018spread},  Zeng et al.~\cite{zeng2016rumors}, Zubiaga et al.~\cite{zubiaga2016analysing}\\

&  \\

\multirow{2}{*}{Facebook} & 
Beutel et al.~\cite{beutel2013copycatch}, Del et al.~\cite{del2016spreading}, Friggeri et al.~\cite{friggeri2014rumor}, Nguyen et al.~\cite{nguyen2012containment},\\
& Silverman et al.~\cite{silverman2015lies,silverman2016analysis}, Tacchini et al.~\cite{tacchini2017some}\\

& \\

\multirow{4}{*}{Review platforms} & 
Akoglu et al.~\cite{akoglu2013opinion}, Beutel et al.~\cite{beutel2014cobafi}, Harris et al.~\cite{harris2012detecting}, Hooi et al.~\cite{hooi2016birdnest}, Jindal et al.~\cite{jindal2008opinion}, \\
& Kumar et al.~\cite{kumar2018rev2}, Li et al.~\cite{li2017bimodal,li2011learning,li2014towards,li2015analyzing}, Lin et al.~\cite{lin2014towards}, Luca et al.~\cite{luca2016fake}, Minnich et al.~\cite{minnich2015trueview}, \\
& Mukherjee et al.~\cite{mukherjee2012spotting,mukherjee2013yelp}, Ott et al.~\cite{ott2011finding,ott2013negative}, Rayana et al.~\cite{rayana2015collective}, Sandulescu et al.~\cite{sandulescu2015detecting}, \\
& Shah et al.~\cite{shah2016edgecentric}, Wang et al.~\cite{wang2011review}, Xie et al.~\cite{xie2012review}, Yao et al.~\cite{yao2017automated}, Ye et al.~\cite{ye2016temporal}\\

&  \\

Sina Weibo & Jiang et al.~\cite{jiang2014inferring},  Kim et al.~\cite{kim2018leveraging},  Ruchansky et al.~\cite{ruchansky2017csi},  Wu et al.~\cite{wu2015false}, Yang et al.~\cite{yang2012automatic}\\

&  \\

Multi-platform & Reddit+Twitter+4chan: Zannettou et al.~\cite{zannettou2017web} \\

&  \\

\multirow{4}{*}{Other} & Fake news articles: Horne et al.~\cite{horne2017fake},  Silverman et al.~\cite{silverman2016analysis},  Rubin et al.~\cite{rubin2016fake},  Perez et al.~\cite{perez2017automatic},  \\
& Wikipedia: Kumar et al.~\cite{kumar2016disinformation}, False information websites: Albright et al.~\cite{albrightelection,albrightdata}, \\
& Fact checking website: Shu et al.~\cite{shu2017exploiting} and Wang et al.~\cite{wang2017liar}, \\
&  Crowdsourcing and tabloid websites: Perez et al.~\cite{perez2017automatic}\\
\hline
\end{tabular}
\end{center}
\caption{This table categorizes research on false information based on the platforms they study. \label{tab:fakeplatforms}}
\end{table}

Finally, several algorithms have been created for effective \textbf{\textit{detection of false information}} from its true counterparts.
These algorithms can broadly be categorized into three categories: feature-based, graph-based, and propagation-modeling based.
Feature-based algorithms leverage the unique characteristics for detection by using them as features in a machine learning model or rule-based framework~\cite{gupta2013faking,horne2017fake,jindal2008opinion,kumar2016disinformation,ott2011finding,qazvinian2011rumor,sandulescu2015detecting}.
Graph-based algorithms are developed to identify dense-blocks or dense subgraphs of users and information in the network~\cite{akoglu2010oddball,beutel2013copycatch,jiang2014inferring,kumar2018rev2,rayana2015collective,wang2011review}.
Propagation-modeling algorithms create information spread models for true information and use these models to identify false information~\cite{acemoglu2010spread,budak2011limiting,jin2013epidemiological,nguyen2012containment,tripathy2010study,yang2012automatic,wu2015false}.
Naturally, the accuracy of these algorithms depends on the task and datasets used.
However, several reach the high eighties and nineties, showing their effectiveness on large-scale real-world datasets of fake reviews, fake news, and hoaxes.
Detection algorithms are discussed in depth in Section~\ref{sec:detection}.

Overall, this survey gives a comprehensive overview of the state of false information on the web and social media. Table~\ref{tab:fakeplatforms} categorizes the research papers according to the platforms they study.

The remainder of the survey is organized as follows: Section~\ref{sec:type} explains the two broad categories of false information, Section~\ref{sec:rationale} discusses the mechanisms and rationale for the success of false information. Then, in  Section~\ref{sec:impact}, we describe the impact of false information. Section~\ref{sec:characteristics} elaborates on various characteristics of false information, and is followed by Section~\ref{sec:detection} which describes several algorithms for its detection.

\section{\textbf{Types of false information}}
\label{sec:type}

\begin{figure}
\centering
\includegraphics[width=\columnwidth]{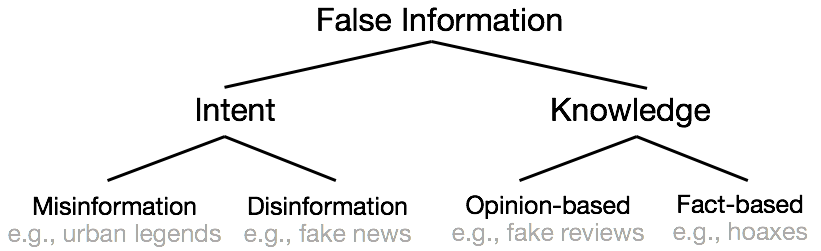}
\caption{Categorization of false information based on intent (i.e., is it spread with the intention to deceive or not) and knowledge (i.e., if there is a single ground  \label{fig:categorization}}

\end{figure}

False information can be categorized based on its intent and knowledge content, as depicted in Figure~\ref{fig:categorization}. We discuss and detail this categorization here.

\subsection{\textbf{Categorization based on intent}}
False information can be classified based on the intent of the author, as \textit{misinformation} and \textit{disinformation}~\cite{fallis2014functional, hernon1995disinformation}.
By definition, misinformation is spread \emph{without} the intent to deceive.  Thus, common causes of misinformation include misrepresentation or distortion of an original piece of true information by an actor, due to lack of understanding, attention or even cognitive biases~\cite{fallis2009conceptual,skyrms2010signals}.
These actors can then spread misinformation unwittingly to others via blogs, articles, comments, tweets, and so on.
Note that readers can often simply have different interpretations and perception of the same piece of true information, leading to differences in how they communicate their understandings and in turn inform others' perception of the facts~\cite{aikin2009poe, krech1948perceiving}.

Conversely, disinformation is spread \emph{with} the intent to deceive~\cite{pomerantsev2014menace}. Thus, understanding the motives for disinformation are much akin to understanding the motives for deception~\cite{fallis2014functional,skyrms2010signals}. Deception on the web occurs for many purposes, and for similar (though less interpersonal) reasons as in human interactions. The large potential audience leads most web disinformation campaigns to focus on swaying public opinion in one way or another, or driving online traffic to target websites to earn money by advertisements. One recent example is political disinformation spread during 2016 USA presidential elections~\cite{forelle2015political, howard2016bots, shao2017spread}, which even led to public shootings~\cite{fisher2016pizzagate}.
In this survey, we focus primarily on the technological aspects of disinformation, as the majority of research focuses around it.

\subsection{\textbf{Categorization based on knowledge}}
Under this categorization, false information is classified as either \textit{opinion-based} or \textit{fact-based}~\cite{thomas1986statements}.
Opinion-based false information expresses individual opinion (whether honestly expressed or not) and describes cases in which there is no absolute ground truth.
The creator of the opinion piece knowingly or unknowingly creates false opinions, potentially to influence the readers' opinion or decision.
An example of false information that lies in this category is fake reviews of products on e-commerce websites, where people express their opinions about product quality.
On the other hand, fact-based false information involves information which contradicts, fabricates, or conflates a single-valued ground truth information.
The motive of this type of information is to make it harder for the reader to distinguish true from false information, and make them believe in the false version of the information~\cite{pomerantsev2014menace}.
This type of false information includes fake news, rumors, and fabricated hoaxes.
There is significant research in both opinion-based and fact-based false information, and we will discuss both in this survey.

\begin{figure}
\centering
\includegraphics[width=0.6\textwidth]{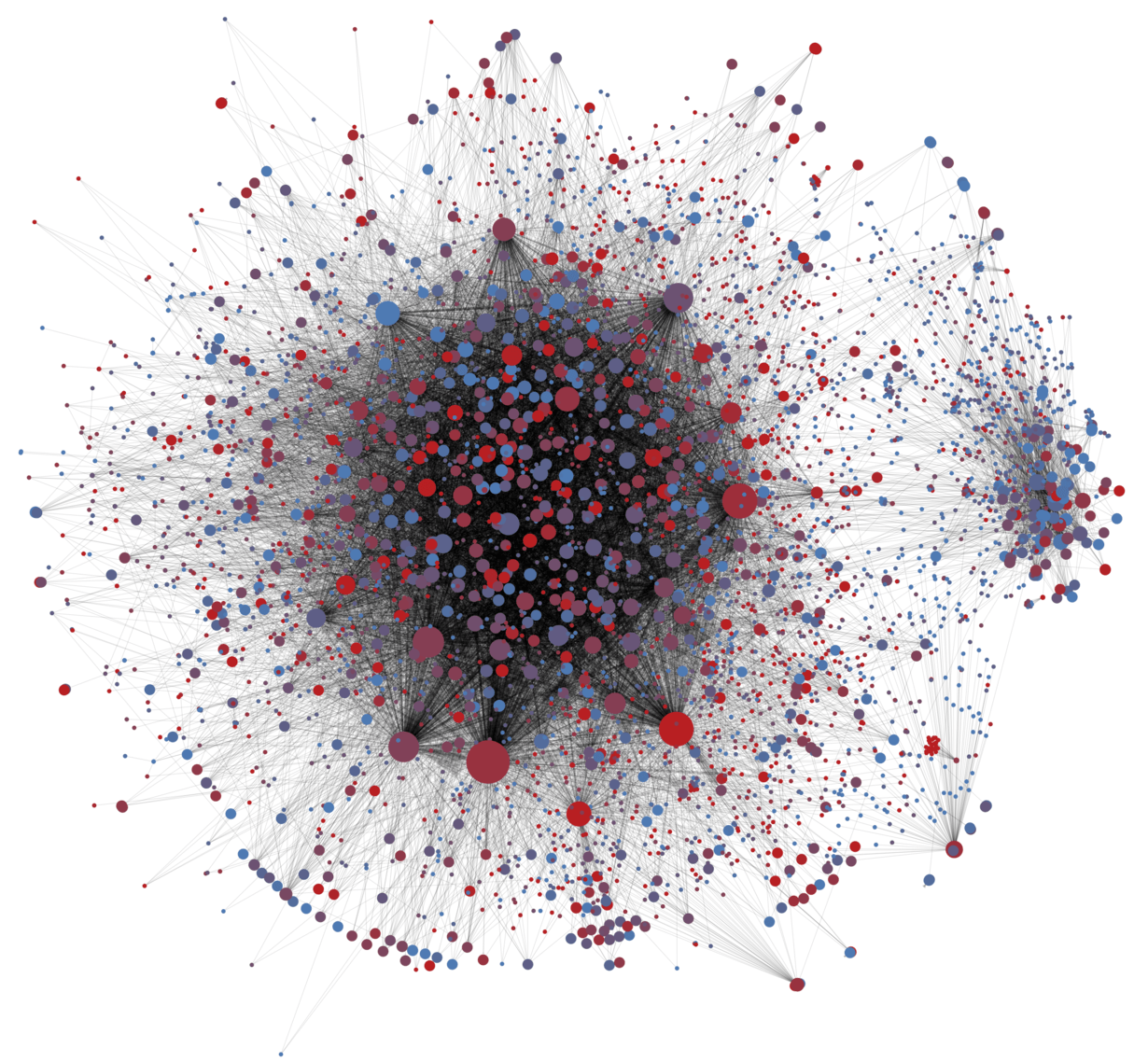}
\caption{\label{fig:spread_fake_news}False information spreads in social media via bots: Figure shows spread of \#SB277 hashtag concerning a vaccination law. Red dots are likely bots and blue are likely humans. Figure reprinted with permission from~\cite{truthypage}.}\footnote{For this and all subsequent reprinted figures, the original author(s) retain their copyrights, and permission was obtained from the author(s).}
\end{figure}

\section{\textbf{Actors and rationale of successful deception by false information}}
\label{sec:mechrat}

This section describes the types of mechanisms used for spreading false information and the rationale behind their success. 

\subsection{\textbf{Bad actors: bots and sockpuppets}}
\label{sec:mechanisms}
Humans are susceptible to false information and spread false information~\cite{vosoughi2018spread}.
However, the creation and spread of false information is complex, and fueled by the use of nefarious actors which act independently or on a large-scale using a network of social media bots.
Both deceive readers by creating an illusion of consensus towards the false piece of information, for instance by echoing it multiple times or expressing direct support for it.
These accounts aim to artificially \textit{engineer} the virality of their content (e.g., by `upvoting'/promoting content in its early phase~\cite{weninger2015random}) in order to spread posts with false information even faster and deeper than true information~\cite{bakshy2011everyone,cheng2014can,friggeri2014rumor}.

Lone-wolves operate by creating a handful of fake ``sockpuppet'' or ``sybil'' accounts and using them in coordination to reflect the same point of view, by writing similar reviews on e-commerce platforms or making similar comments on public forums.
Lone-wolf operations using multiple accounts can be especially convincing as readers are typically not aware that a whole discussion is fabricated and actually originates from a single source.
For instance, in online conversations, Kumar et al.~\cite{kumar2017army} characterize this behavior by studying 61 million comments made by 2.1 million users across several discussion platforms.
They found that while sockpuppets can be used with benign intention, sockpuppets with deceptive intentions are twice as common.
Deceptive sockpuppets reply to each other with agreement and support, and are negative towards accounts that disagree.
Moreover, these accounts hold central locations in the communication network, and are therefore in key spots to spread false content.
Similarly, sybil accounts in communication and social networks are created to integrate themselves well into the network and prevent detection in order to increase their influence over others~\cite{yang2014uncovering}.

On a larger scale, social media botnets are used to spread false information.
Bots, which are fake or compromised accounts controlled by a single individual or a program, are used to serve two main purposes: to send the same information to a large audience quickly, and to inflate the ``social status'' of certain users, both of which make false information to appear credible and legitimate~\cite{ferrara2016rise,shah2017manyfaces,subrahmanian2016darpa}.
Figure~\ref{fig:spread_fake_news} visualizes an online Twitter conversation on a controversial topic (hashtag \#SB277) showing overwhelming presence of bots (red nodes) engaging with humans (blue nodes)~\cite{truthypage}.
Bessi et al.~\cite{bessi2016social} and Shao et al.~\cite{shao2017spread} studied the use of bots in political campaigns and found that bot accounts are responsible for almost one-fifth of all Twitter political chatter, and that false information is more likely to be spread by bots than real users.
Similarly, Nied et  al.~\cite{nied2017alternative} found that 25\% of false information tweets were generated by bots.
A common strategy employed by bots is to target information towards more influential real users, who may sometimes get influenced and reshare the false message forward to a broader audience~\cite{bessi2016social,shao2017spread}.
In efforts to increase ``social status'', botnet operators offer services that provide fake followers by using their bots to follow paying customer accounts.
Shah et al.~\cite{shah2017manyfaces} studied these services and found that they operate on ``freemium'' and ``premium'' models, where the former is made of compromised or real user accounts and the latter is comprised of fake or bot accounts. These two models operate quite distinctly---freemium fraud accounts create high-density cliques of opted-in accounts who trade follows amongst themselves, while premium fraud accounts create dense bipartite cores, i.e., one set of accounts follows the paying customers. This increases the apparent trustworthiness of the users, who can then be used to spread false information further.

In a recent study, Vosoughi et al.~\cite{vosoughi2018spread} analyzed over 126,000 false information cascades on Twitter over a period of 11 years and showed that humans were responsible for spread of false information on Twitter, not bots.
Using the BotOrNot Twitter bot-detection tool developed by Davis et al~\cite{davis2016botornot}, they identified the bot and non-bot accounts that engaged in false information spread.
They found that on Twitter, humans, not bots, were responsible for spread of false information, as the bots were responsible for accelerating the spread of both true and false information roughly equally.
Even after removing bot activity, false information was observed to spread farther, deeper, faster, and broader than true information.
Further, they found that the non-bot accounts on Twitter that were responsible for spreading false information were newer, had fewer followers and followees, and were less active.
While this is the case on Twitter, other platforms may behave differently, and the proliferation of nefarious actors in the creation and spread of false information is common.

Thus, using sockpuppets and botnets are used to engineer the spread of false information to massive numbers of real users on social media. These accounts operate using fake and computerized accounts to increase the visibility of false information and social status of accounts that spread it.

\subsection{\textbf{Rationale of successful deception by false information}}
\label{sec:rationale}
In the previous section, we discussed the multifaceted motives and spreading mechanisms used by those who publish false information. But what about the susceptibility of its consumers: do the readers tend to believe it, and if so, why?

\begin{figure}
\centering
\includegraphics[width=\textwidth]{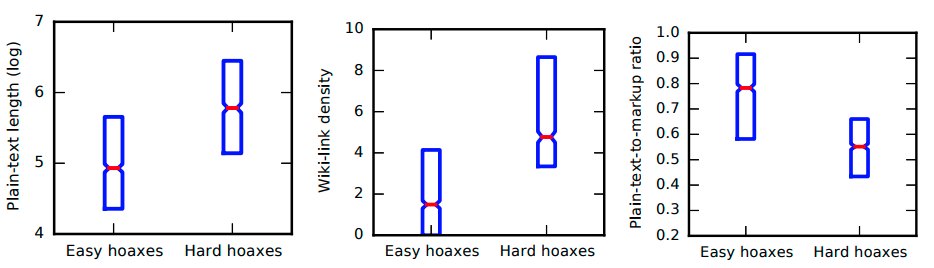}
\caption{Humans are unable to identify false information that is crafted to look genuine, as demonstrated in this case of hoaxes on Wikipedia. Reprinted with permission from~\cite{kumar2016disinformation}.\label{fig:susceptible-hoax}}

\end{figure}

\subsubsection{\textbf{Human inability to discern false information}}
False information would not have any influence if readers were able to tell that it is false. 
However, several research studies have conducted experiments to measure the ability of humans to detect false information including hoaxes, fake reviews, and fake news, and have shown that humans are not particularly good at discerning false from true information~\cite{kumar2016disinformation,ott2011finding,perez2017automatic,yao2017automated}.
We describe these studies in detail below.

To understand reader susceptibility, Kumar et al.~\cite{kumar2016disinformation} conducted an experiment with hoax articles created by hoaxsters on Wikipedia.
They hired Amazon Mechanical Turk workers and showed them one hoax and one non-hoax article side-by-side, with the task to identify which one of the two articles was a hoax article without searching for information elsewhere.
A total of 320 pairs of hoax and non-hoax articles were created and each pair was shown to 5 different workers.
Humans correctly identified the hoax a mere 66\% of times, only marginally higher than the random guessing baseline of 50\%.
They further studied the reasons for mistakes that workers made, shown in Figure~\ref{fig:susceptible-hoax}, which compares several statistical properties of easily-identifiable and hard-to-identify hoaxes.
They found that workers frequently misjudged long and well referenced hoax articles to be true, and short but true articles that lacked references to be hoaxes.
In fact, even trained and trusted Wikipedia volunteers, called ``patrollers,'' make the similar mistakes
by approving long and well-referenced hoax articles for publication on Wikipedia instead of rejecting and deleting them.
So, \textit{if false information is purposefully created to look genuine, both trained and casual readers are deceived}.
This indicates that humans give a lot of emphasis on the appearance of false information when judging its veracity.

In the domain of fake reviews, several of research studies have come to similar conclusions.
Ott et al.~\cite{ott2011finding} demonstrated that humans are not very good at discerning deceptive opinion spam from real reviews. As a compelling example, below are two TripAdvisor reviews (one real and one fake). Can you identify which one is fake?\footnote{The second review is fake.}

\begin{quote}
\textit{``I have stayed at many hotels traveling for both business and pleasure and I can honestly stay that The James is tops. The service at the hotel is first class. The rooms are modern and very comfortable. The location is perfect within walking distance to all of the great sights and restaurants. Highly recommend to both business travelers and couples.''}
\end{quote}

\begin{quote}
\textit{``My husband and I stayed at the James Chicago Hotel for our anniversary. This place is fantastic! We knew as soon as we arrived we made the right choice! The rooms are BEAUTIFUL and the staff very attentive and wonderful!! The area of the hotel is great, since I love to shop I couldn't ask for more!! We will definatly [sic] be back to Chicago and we will for sure be back to the James Chicago.''}
\end{quote}

The fake reviews were generated by Amazon Mechanical Turkers.
Three humans were given a total of 160 reviews which contained both real and fake reviews, and workers had an accuracy between 53.1\% and 61.9\% in identifying the fake reviews, again showing that humans are poor judges of deception, and perform close to random.

For fake news, a similar recent study was conducted by Perez et al.~\cite{perez2017automatic}.
They created a dataset of crowdsourced and crawled celebrity-oriented real and fake news, and gave 680 pieces of news (50\% fake) to two humans to identify fake ones from them.
They achieved an average accuracy of 70.5\% in detecting made-up crowdsourced news, and 78.5\% in detecting celebrity news. 

More recently, with the advancement in deep learning, false information can be generated automatically. When fine tuned, this false information can be as deceptive as those created by humans.
Yao et al.~\cite{yao2017automated} created a deep neural network model that generates fake reviews for restaurants, by training on Yelp review data. Mechanical Turk workers were shown a set of 20 reviews for each restaurant, which contained between 0 and 5 machine generated reviews. The task of the workers was to identify which of the reviews were fake.
A total of 600 sets of reviews were labeled, and the workers achieved a very low precision of 40.6\% precision and 16.2\% recall.
Humans were able to identify fake reviews if they contained repetitive errors, but not when they had minor spelling or grammar mistakes.

Altogether, these four studies show that humans can easily be deceived into believing that false information is true when it is created intelligently to appear like true information, both manually or by machines.

\begin{figure}
\centering

\includegraphics[width=0.6\textwidth, height=2.5in]{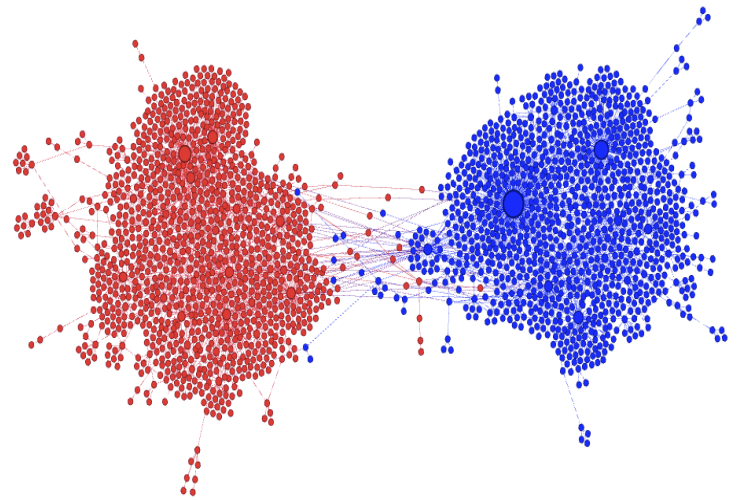}

\caption{Social media platforms can produce echo-chambers, which lead to polarization and can encourage the spread of false information. Figure shown echo-chambers formation in retweet graph on controversial \#beefban topoc. Reprinted with permission from~\cite{garimella2017balancing}.\label{fig:echochamber}}

\end{figure}

\subsubsection{\textbf{Formation of echo-chambers}}
Given the advent of improved recommendation algorithms which promote personalized content for easy user access and exposure,  social media platforms are often party to an ``echo-chamber'' effect~\cite{del2016spreading}. This effect primarily refers to the self-selective polarizing effect of content where people immerse themselves in social circles in such a way that they are primarily exposed to content that agree with their beliefs.
For example, a political liberal might friend more liberals on Facebook, thumbs-up liberal-minded content, and thus constantly be exposed to posts and news which aligns with his worldview.
Figure~\ref{fig:echochamber} visualizes this echo-chamber effect on Twitter on a controversial topic of \#beefban, where red and blue nodes represent users with opposing beliefs and edges represent who-retweets-whom, as shown by Garimella et al.~\cite{garimella2017balancing}.
Notice that both groups are mostly disconnected with few messages between nodes of different types.
The echo-chamber effect in social networks is substantiated by Nikolov et al.~\cite{nikolov2015measuring} by demonstrating that the diversity of sources (links) clicked by users is significantly lower on social media platforms than in general search engines.
Several studies have studied the effects and causes of echo-chambers.
Quattrociocchi et al.~\cite{quattrociocchi2016echo} demonstrated that such resulting echo-chambers can serve to polarize the user's viewpoints by means of confirmation bias and lead to less diverse exposure and discussion between unaligned users.  The resulting echo-chambers can contribute to the spread of false information by lowering the bar for critical fact-checking. 
Moreover, Trilling et al.~\cite{trilling2017exposure} and Zajonc \cite{zajonc1968attitudinal} posited that the perceived accuracy of false information increases linearly with the frequency of exposure of a participant to the same false information. This suggests that familiarity with repeatedly shared content (highly common and expected in echo-chambers) increases the perceived accuracy of the content, irrespective of its credibility.
This calls for research on how to create effective techniques to break echo-chambers and slow down false information spread.

\subsubsection{\textbf{Other reasons of successful deception}}
Publishers of false information succeed at deceiving and spreading it by playing upon naiveté and biases of consumers.
Flynn et al.~\cite{flynn2017nature} showed that prior belief in false information is rooted in the biased reasoning of the presented information. Two major factors that make consumers vulnerable or susceptible to believing false information are \emph{confirmation bias} and \emph{naive realism}~\cite{shu2017fake}.
Naive realism suggests consumers believe that they have the ``true'' perception of reality whereas disagreements or nonalignment of views is construed as the others' lack of rationality or cognizance~\cite{ward1997naive}.
Moreover, Nickerson et al.~\cite{nickerson1998confirmation} characterized \textit{confirmation bias}, or the tendency of consumers to seek or interpret evidence which confirms their pre-existing notions or beliefs.
These biases lead consumers to look for and find meaning in pieces of information (no matter the veracity) which substantiate their own claims.
For example, political liberals are prone to having more affinity towards posts promoting liberal viewpoints and condemning conservative ones, and vice versa.
Furthermore, \textit{social normative theory} suggests that sharing content aligned with the beliefs of their peers is attractive~\cite{asch1951effects}, in order to gain the acceptance or favor of their peers, regardless of its veracity.
Alarmingly, Nyhan et al.~\cite{nyhan2010corrections} showed that even the presentation of corrections to false information by means of facts can actually further polarize idealogical groups and \emph{increase} their misperceptions.

Researchers have explored the psychological and demographics of information consumers and their propensity to believe in it.
Pennycook et al.~\cite{pennycook2017whofalls} investigated the psychological profiles to show a positive correlation between propensity for analytic thinking and the ability to discern false from real information, suggesting that false information often spreads due to poor analytical skills on the part of the consumer spreader.
Additionally, inability to discern the publisher's original (possibly genuine) intentions can lead to consumer's misunderstanding of original information and lead to creation of misinformation~\cite{aikin2009poe}.
Recent analysis of demographics by Allcott and Gentzkow~\cite{allcott2017social} concluded that people who spend more time consuming media, people with higher education, and older people have a more accurate perception of information.

\vspace{2mm}
Overall, false information spread is orchestrated on large-scale by using fake social media accounts, such as bots and sockpuppets. Once their false message spreads and deceives readers, the readers themselves echo and spread the message. Several reasons lead to deception by false information. First, humans are unable to distinguish false information from true ones when they come across them, and this is difficult even when they are trained to do so. Second, echo chambers are formed in social platforms, such that true and false information is spread among different groups of users, which do not interact with one another. And finally, human biases lead to increase in susceptibility, with some demographics (less educated and low consumers of media) being more likely to fall for false information.

\section{\textbf{Impact of false information}}
\label{sec:impact}
Given that there are several factors that lead to deception by false information (Section~\ref{sec:mechanisms}), what is the impact of false information on its readers on web and social media?
In the real world, false information has been shown to have significant impact on the stock market~\cite{bollen2011twitter}, hampering response during natural disasters~\cite{gupta2013faking}, and terroristic activity~\cite{fisher2016pizzagate,starbird2014rumors}.
On web and social media, the impact is measured as the engagement it produces via its readers, using statistics such as number of reads, number of days it survived without being removed, or number of people it reached via reshares.
Several research studies have been conducted to measure the impact of hoaxes, fake reviews, and fake news.
Frigerri et al.~\cite{friggeri2014rumor} studied the spread of rumors on Facebook, Kumar et al.~\cite{kumar2016disinformation} measured the impact of hoax articles on Wikipedia, and Silverman~\cite{silverman2016analysis} analyzed the engagement of fake election news articles on Facebook.
We discuss these studies to measure the impact of false information on web platforms.

\begin{figure}
\centering
\includegraphics[width=0.3\textwidth]{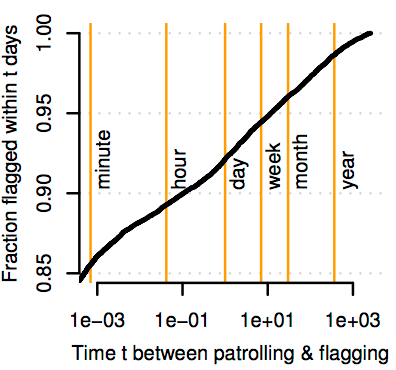}
\includegraphics[width=0.3\textwidth]{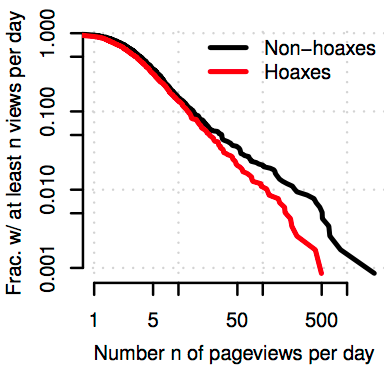}
\caption{False information is impactful as it (a) survives for a long time, and (b) is viewed by thousands of people. Reprinted with permission from \cite{kumar2016disinformation}. \label{fig:impact_hoax}}

\end{figure}

False information spreads far and wide on social media because there is an average delay of 12 hours between start of false information spread and that of its debunking information~\cite{zubiaga2016analysing, shao2016hoaxy}.
False information spreads rapidly during its starting phase---an unverified and not yet debunked rumor has high potential of becoming viral~\cite{zubiaga2016analysing}.
As a result, rumors with the possibility of being true start to spread, sometimes even by reputed news organizations~\cite{silverman2015lies}.

On Wikipedia, Kumar et al.~\cite{kumar2016disinformation} measured impact of hoaxes in terms of their viewcount, number of days they survived before they are deleted, and their spread across the web. Figure~\ref{fig:impact_hoax} shows the distributions for the first two statistics. Figure~\ref{fig:impact_hoax}(a) shows the distribution of the time it takes from when the article is created and approved (`patrolled') till the time it is identified as a hoax (`flagged'). It shows that while 90\% of hoax articles are identified immediately within an hour of being approved, about 1\% of hoaxes that are well-written hoaxes survive for over one year without being detected.
However, survival is not enough for a hoax article to be successful; it must be viewed as well.
Figure~\ref{fig:impact_hoax}(b) plots the counter-cumulative distribution of average view count of hoaxes that survive for at least a week and their equivalent non-hoaxes.
On average, hoaxes are viewed less frequently than non-hoaxes (median 3 views per day vs 3.5 views per day), but a non-negligible 1\% of hoaxes are viewed at least 100 times a day. Finally, the impact of hoaxes is measured in terms of spread over the web, by counting the links that were clicked by readers to reach the hoax article. For this, 5 months of Wikipedia server logs were used. The results were alarming---at least 5 distinct links were clicked from across the web for 7\% of hoaxes, and on average, each hoax had 1.1 such links. This traffic was observed from search engines, social networks such as Facebook and Twitter, and from within Wikipedia itself. Overall, this analysis shows that while most hoax articles are ineffective, a small fraction of hoax articles on Wikipedia is highly impactful.

Buzzfeed news analyzed highly impactful fake political news on the web. They analyzed both true and false election-related stories with the highest engagement on Facebook during the 2016 US Presidential election~\cite{silverman2016analysis}. Engagement was measured as the total number of shares, reactions, and comments on the Facebook story. They analyzed the 20 top-performing false election stories generated by fake websites and blogs, and compared them to the 20 top-performing true election stories from major news websites, like New York Times, Washington Post, and others. The fake news stories got a total of 8,711,000 engagements, significantly higher than the 7,367,000 engagements of the real news stories.
As this analysis was restricted to top stories, a complete analysis of all news stories may reveal a different picture.
Prior to this study, Gupta et al.~\cite{gupta2013faking} studied the spread of eight fake images on Twitter during Hurricane Sandy, and found that that fake images were shared almost twice as much as real images.

On a larger scale, Frigerri et al.~\cite{friggeri2014rumor} conducted a comprehensive study of the spread of false and real information on Facebook.
They collected 4,761 rumors from \emph{snopes.com}, which is a website that catalogues popular stories on social media and checks their veracity.
In their dataset, 45\% of stories were fake, 26\% were ``true'' (i.e., not a fake story), and the rest had intermediate truth values.
They analyzed the rumor cascades propagating as photos on Facebook during July and August 2013. Each cascade was identified as a tree of reshares starting from the original post of the photo, whenever a link to a valid snopes article was posted as a comment to the original photo or one of its reshares. A total of 16,672 such cascades were identified, with 62,497,651 shares, showing the large visibility false rumors can have. 
Surprisingly, they found that false information cascades were deeper, as there were more reshares at greater depths than the reference cascades. At lower depth, i.e., closer to the original photo post, the reference cascades have more reshares---about 20\% reference cascades have depth of at least two, compared to 10\% of false information cascades. But the reference cascades die very soon, while false information cascades run deeper. About 3\% of false cascades have depth of at least 10 reshares, while less than 1\% of reference cascades have the same property. The difference increases in magnitude as the depth of the cascade increases. This study shows the large reach of false information on social media, fueled by its highly contagious nature.

Recently, the largest study of spread of over 126,000 rumors on Twitter over a period of 11 years was conducted by Vosoughi et al.~\cite{vosoughi2018spread}.
The authors took the set of false information cascade identified by various independent fact-checking agencies and traced their spread from their very beginning.
This was done by identifying cascades that contained a link to any of the agencies.
For comparison, they also considered cascades of verified true information linking to these agencies.
Compared to true information, tweets containing false information spread significantly farther (more number of users retweeted), faster (more number of retweets in a shorter time), deeper (more number of retweet hops), and more broadly (more number of users at some retweet depth).
This was observed in all categories of false information, such as politics, urban legend, science, business, and others, with politics as the biggest category of false information.
In fact, they found that the top 1\% of false tweets reached over 1,000 users, which true information tweets rarely did.
False information reached more number of people than truth at every cascade depth, which was aided by its virality, showing that it was spread by multiple people in a peer-to-peer manner, instead of a few accounts simply broadcasting it.
Moreover, false information was six times faster in reaching the same number of people as true information did.
Thus, this study showed the widespread reach and impact of false information in Twittersphere.

Overall, impact of false information on the web is measured using engagement statistics such as view count, share count, and more.
Research has shown that while most false information is not effective, a small fraction (typically 1\%) is highly impactful, most popular false information pieces attract more attention than real information, and false information spreads widely and quickly across the web and reaches a large population on social media.

\begin{table}
\begin{tabular}{|c|c|c|}
\hline
\textbf{Feature} & \textbf{Opinion-based false information} & \textbf{Fact-based false information}\\
\textbf{category} &  \textbf{(fake reviews)}  &  \textbf{(false news and hoaxes)}  \\
\hline

\multirow{6}{*}{Text} &  
Harris et al.~\cite{harris2012detecting}, Jindal et al.~\cite{jindal2008opinion},  & 
Gupta et al.~\cite{gupta2013faking},  Horne et al.~\cite{horne2017fake},  \\

& Li et al.~\cite{li2011learning} , Lin et al.~\cite{lin2014towards}, &   
Howard et al.~\cite{howard2016bots}, Kumar et al.~\cite{kumar2016disinformation},  \\ 

&  Mukherjee et al.~\cite{mukherjee2012spotting,mukherjee2013yelp} , Ott et al.~\cite{ott2011finding,ott2013negative},  & Mitra et al.~\cite{mitra2015credbank,mitra2017parsimonious}, Perez et al.~\cite{perez2017automatic},   \\

& Rayana et al.~\cite{rayana2015collective}, Yao et al.~\cite{yao2017automated} & Qazvinian et al.~\cite{qazvinian2011rumor}, Rubin et al.~\cite{rubin2016fake},  \\
& & Silverman et al.~\cite{silverman2015lies}, Wang et al.~\cite{wang2017liar}, \\
&& Yang et al.~\cite{yang2012automatic}, Zeng et al.~\cite{zeng2016rumors}\\

&& \\

\multirow{4}{*}{User} & 
Hooi et al.~\cite{hooi2016birdnest},  Kumar et al.~\cite{kumar2018rev2}, & Bessi et al.~\cite{bessi2016social},  Davis et al.~\cite{davis2016botornot},  Gupta et al.~\cite{gupta2013faking} \\

&  Li et al.~\cite{li2011learning}, Minnich et al.~\cite{minnich2015trueview}, &  Jin et al.~\cite{jin2013epidemiological}, Kumar et al.~\cite{kumar2016disinformation},  Mendoza et al.~\cite{mendoza2010twitter} \\

& Mukherjee et al.~\cite{mukherjee2012spotting}, Rayana et al.~\cite{rayana2015collective}, & Nied et al.~\cite{nied2017alternative},  Shao et al.~\cite{shao2016hoaxy,shao2017spread} , \\
& Shah et al.~\cite{shah2016edgecentric}   & Tacchini et al.~\cite{tacchini2017some}, Vosoughi et al.~\cite{vosoughi2018spread},   \\
& & Yang et al.~\cite{yang2012automatic}\\
&& \\

\multirow{4}{*}{Graph} & 
Lin et al.~\cite{lin2014towards},  Minnich et al.~\cite{minnich2015trueview}, & 
Bessi et al.~\cite{bessi2016social},  Davis et al.~\cite{davis2016botornot}, Friggeri et al.~\cite{friggeri2014rumor}, \\

& Pandit et al.~\cite{pandit2007netprobe}, Rayana et al.~\cite{rayana2015collective},  &  Kumar et al.~\cite{kumar2016disinformation}, Mendoza et al.~\cite{mendoza2010twitter},  Nied et al.~\cite{nied2017alternative},\\

& Shah et al.~\cite{shah2017flock} &   Qazvinian et al.~\cite{qazvinian2011rumor}, Starbird et al.~\cite{starbird2017examining} \\
&& Subrahmanian et al.~\cite{subrahmanian2016darpa},  Vosoughi et al.~\cite{vosoughi2018spread} \\

&& \\

& Beutel et al.~\cite{beutel2014cobafi},  Harris et al.~\cite{harris2012detecting},  &  \\
Rating &  Hooi et al.~\cite{hooi2016birdnest}, Kumar et al.~\cite{kumar2018rev2}, Li et al.~\cite{li2011learning},   & \\
score & Luca et al.~\cite{luca2016fake}, Minnich et al.~\cite{minnich2015trueview},   & Not applicable \\
& Mukherjee et al.~\cite{mukherjee2012spotting,mukherjee2013yelp}, Rayana et al.~\cite{rayana2015collective},  & \\
& Shah et al.~\cite{shah2016edgecentric}, Ye et al.~\cite{ye2016temporal} & \\
&& \\

\multirow{4}{*}{Time} & 
Hooi et al.~\cite{hooi2016birdnest},  Li et al.~\cite{li2015analyzing,li2017bimodal},   & 
Davis et al.~\cite{davis2016botornot},  Del et al.~\cite{del2016spreading},  Friggeri et al.~\cite{friggeri2014rumor} \\

& Minnich et al.~\cite{minnich2015trueview}, Mukherjee et al.~\cite{mukherjee2012spotting},  &  Shao et al.~\cite{shao2017spread}, Vosoughi et al.~\cite{vosoughi2018spread}, \\

& Rayana et al.~\cite{rayana2015collective}, Shah et al.~\cite{shah2016edgecentric},  & Zannettou et al.~\cite{zannettou2017web}, Zeng et al.~\cite{zeng2016rumors} \\
&Xie et al.~\cite{xie2012review},  Ye et al.~\cite{ye2016temporal} & \\
&& \\

\multirow{4}{*}{Propagation} & & Friggeri et al.~\cite{friggeri2014rumor},  Jin et al.~\cite{jin2013epidemiological},  Shao et al.~\cite{shao2016hoaxy,shao2017spread},  \\
& Not applicable &  Silverman et al.~\cite{silverman2015lies,silverman2016analysis}, Vosoughi et al.~\cite{vosoughi2018spread}, \\
&&  Yang et al.~\cite{yang2012automatic}, Zannettou et al.~\cite{zannettou2017web}, \\
&& Zeng et al.~\cite{zeng2016rumors},  Zubiaga et al.~\cite{zubiaga2016analysing}\\

\hline
\end{tabular}
\caption{This table categorizes research based on the characteristics and features of false information analyzed.\label{tab:chars}}
\end{table}

\section{\textbf{Characteristics of false information}}
\label{sec:characteristics}
In this section, we describe several characteristics of false information that act as ``tell-tale'' signs for discernment from true information. These characteristics are based on textual content, time, ratings, graph structure, creator properties and more. We separately discuss characteristics of opinion-based and fact-based false information, along these axes.
Table~\ref{tab:chars} categorizes the research papers according to the features that they study, as it is one of the most common approaches to approach the problem of false information. The types of features can broadly be grouped as text, user, graph, rating score, time, and propagation-based features. We will discuss these in detail in this section.

\subsection{\textbf{Opinion-based false information}}

Below, we discuss several discovered properties of fake ratings and e-commerce reviews.  We categorize properties pertaining to i) text, ii) ratings, iii) time, iv) graph structure and v) other attributes separately, as these aspects have both individually and in conjunction received considerable attention from researchers.

\begin{figure}
\centering

\includegraphics[width=0.45\textwidth]{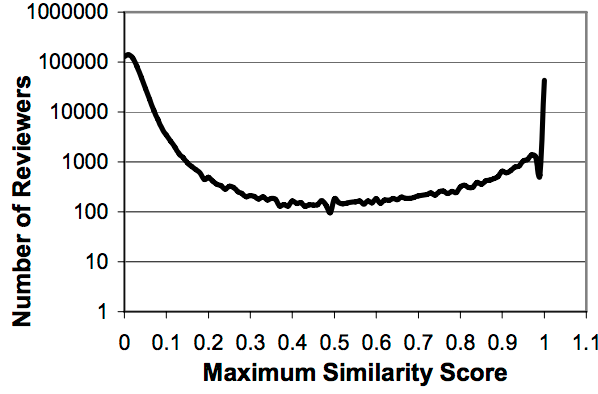}

\caption{Plot showing maximum similarity of reviews between two reviewers, showing that 6\% reviewers have at least one identical review to someone else's review. Reprinted with permission from \cite{jindal2008opinion}.\label{fig:review_similarity}}

\end{figure}

\subsubsection{\textbf{Textual characteristics}}
Since most reviews include textual content, researchers have extensively studied textual and linguistic features for discerning review fraud.  Several works have posited that review fraudsters minimize effort by repeating the same reviews.  Jindal et al.~\cite{jindal2008opinion} provided the first well-known characterizations of review fraud, in which the authors characterized duplicate reviews (according to Jaccard similarity) across Amazon data as cases of fraud.  The authors showed that many of these fraudulent duplicate reviews were from the same user on different products, rather than different users on the same product or different products. Figure ~\ref{fig:review_similarity} shows the distribution of maximum similarity between two reviewers' reviews. At the higher similarity end, 6\% of the reviewers with more than one review have a maximum similarity score of 1, which is a sudden jump indicating that many reviewers copy reviews. Furthermore, Sandulescu et al.~\cite{sandulescu2015detecting} showed that many review fraudsters adjust their reviews slightly so as not to post near or exactly similar reviews and be easily caught---instead, these sophisticated fraudsters tend to post semantically similar text (i.e. instead of duplicating ``the hotel room had an excellent view,'' the fraudster might post ``the hotel room had a superb view'' instead).

Researchers have also focused more on the linguistic features of deceptive reviews, such as using stylistic analysis (number of words, characters, etc.), lexical analysis (number of verbs, nouns, etc.), psycholinguistic analysis (LIWC~\cite{pennebaker2001linguistic}), and sentiment analysis (emotion, sentiment, etc.).
Mukherjee et al.~\cite{mukherjee2013yelp} showed that fake reviews were shorter than real reviews, and Ott et al.~\cite{ott2011finding}  found that imaginative ``faked'' writing is typically more exaggerated and consists of more verbs, adverbs, pronouns and pre-determiners.
Furthermore, deceptive text tends to have an increased focus on aspects external to the venue being reviewed (more emphasis on family, vacation, business, etc.)~\cite{ott2011finding}.
Looking at negative reviews, Ott el at.~\cite{ott2013negative} found that fake negative review writers exaggerate too negatively, including words which communicated negative emotion far more than normal reviews (terrible, disappointed, etc.).  Furthermore, fake reviews eschew the use of pronouns such as ``I,'' perhaps in order to distance themselves from the negative sentiments. Similar observations were made by Li et al.~\cite{li2014towards} on fake reviews generated by domain experts.
Finally, Harris \cite{harris2012detecting} demonstrated that deceptive opinion spam tends to be on average less readable than truthful reviews (measured by Average Readability Index), and is also more polarized and sentimental than those reviews, supporting previous results.

\begin{figure}
\centering

\includegraphics[width=0.45\textwidth]{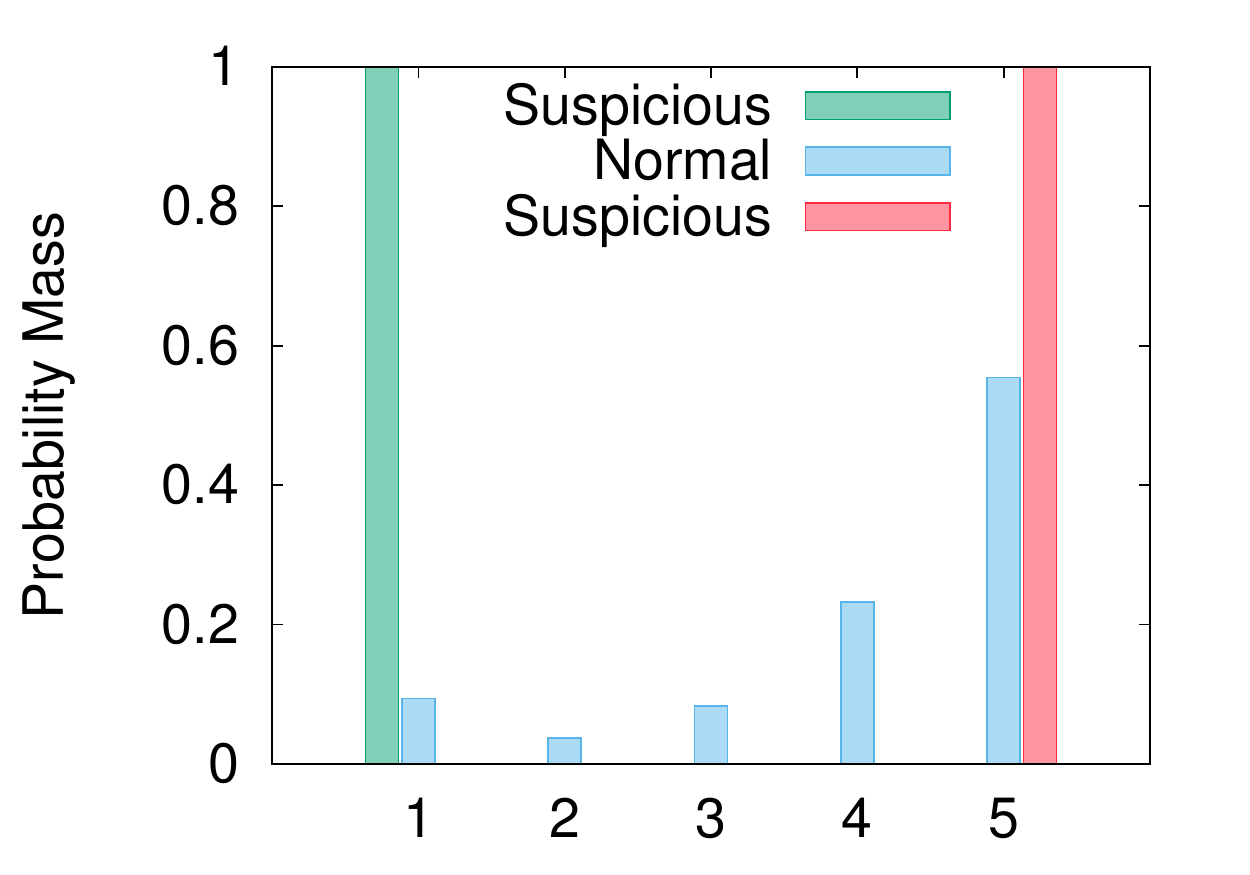}
\includegraphics[width=0.45\textwidth]{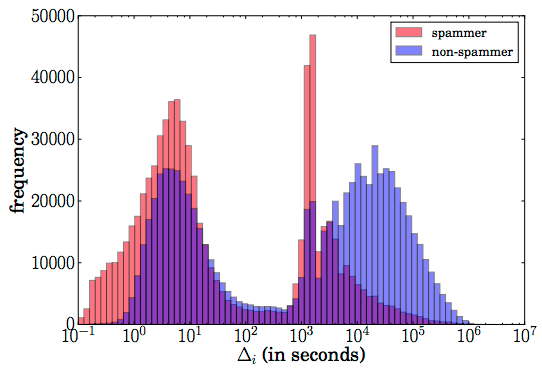}

\caption{(a) Aggregate e-commerce rating behavior typically follows the $J$-shaped curve in blue, whereas  review spammers commonly have strongly positively or negatively-biased rating distributions like those in green and red \cite{shah2016edgecentric}. (b) Fraudulent and non-fraudulent users have bimodal rating distribution~\cite{li2017bimodal}. Figures reprinted with permission from \cite{shah2016edgecentric} and \cite{li2017bimodal}. \label{fig:biased_ratings2}}

\end{figure}

\subsubsection{\textbf{Ratings characteristics} }
Many e-commerce sites disallow users from giving feedback without giving an associated numerical rating.  The rating is typically a 5-star system (1 representing the worst possible rating, and 5 representing the best), and is employed by numerous major online marketplaces including Amazon, eBay, Flipkart, and more.  Prior work in review fraud has shown that those who engage in spreading fake e-commerce reviews also typically have skewed rating distributions \cite{beutel2014cobafi,hooi2016birdnest,rayana2015collective,shah2016edgecentric} which are not typical of real users who share non-uniform opinions over many products.
Figure \ref{fig:biased_ratings2} shows an example from Shah et al.~\cite{shah2016edgecentric}, comparing aggregate (dataset-wide) rating habits from the Flipkart platform with two common, naive fraudster rating habits depicting very positive and negative raters.  Some fraudulent reviewers give only positive ratings as they are created in order to inflate ratings for customer products, whereas other such reviewers give only negative ratings as they intend to slander competitors' products.
Further, Kumar et al.~\cite{kumar2018rev2}
recently showed that fraudulent review writers are typically unfair, in that they give ``unreliable'' rating scores that differ largely from the product's average score. Furthermore, these fraudulent writers often give high ratings to products that otherwise receive highly negative ratings from fair users.

\subsubsection{\textbf{Temporal characteristics}}
Fraudulent review writers typically give reviews in ``lockstep,'' or at the same/similar times.  The rationale is similar to that for dense subgraph connectivity---the review writer's accounts are often controlled by scripts, and are thus temporally synchronized in short windows. A number of papers have leveraged the distribution of interarrival times (IATs) between each user's successive ratings/reviews to detect review spammers. Shah et al.~\cite{shah2016edgecentric}, Hooi et al.~\cite{hooi2016birdnest}, and Ye et al.~\cite{ye2016temporal} showed that in e-commerce websites, spammers are often characteristic of very short IATs (on the order of seconds or minutes) between subsequent ratings, unlike typical users who would rate sporadically and likely only upon making and receiving a purchase. Xie et al.~\cite{xie2012review} substantiated these findings, with particular emphasis on singleton review spammer attacks. Further, Li et al.~\cite{li2015analyzing} and Minnich et al.~\cite{minnich2015trueview} showed that many fraudulent check-ins/reviews in such networks occur with short, but moreover ``infeasible'' IATs between check-ins.  Since users must be physically present at a location to be allowed to check-in and leave a review for a location, reviews at far-away places in very short IATs are a notable distinguishing characteristic of fraud.

In addition, more recent research by Li et al.~\cite{li2017bimodal} surprisingly found that the posting rates of both fraudulent and non-fraudulent users is bimodal---some reviews are written in a short time bursts, while some others with more time between consecutive reviews. Fraudulent users are still more bursty than non-fraudulent users, as the latter have the tendency to be more active after a period of inaction to summarize their recent experiences.

\begin{figure}
\centering
\includegraphics[width=\textwidth]{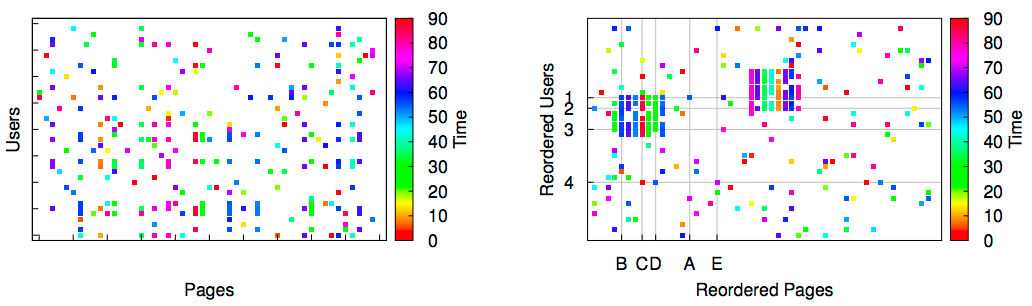}
\caption{Fraudulent reviewers often operate in coordinated or ``lock-step'' manner, which can be represented as temporally coherent dense blocks in the underlying graph adjacency matrix. Reprinted with permission from ~\cite{beutel2016user}.\label{fig:copycatch}}

\end{figure}

\subsubsection{\textbf{Graph-based characteristics}}
Several works show that dense subgraphs produced by coordinated or ``lock-step'' behavior in the underlying connections of the social (in this case, review) graph are associated with fraudulent behavior \cite{beutel2013copycatch, pandit2007netprobe, shah2017flock}. Figure ~\ref{fig:copycatch} demonstrates this pattern in page-likes on Facebook~\cite{beutel2013copycatch}. Alternatively, other works look at the local network structure of the users instead of global structure. For example, Lin et al.~\cite{lin2014towards} showed that for review platforms where multiple ratings/reviews can be given to the same product, review fraudsters often repeatedly post to the same product instead of diversifying like a real reviewer.

Studying group structure, Mukherjee et al.~\cite{mukherjee2012spotting} showed that ratios for review group (defined as a set of reviewers who have reviewed at least $k$ common products) size to the total number of reviewers for an associated product tend to be significantly higher in fraudster groups than real users.  This is because many products (especially bad ones) have ratings/reviews almost entirely given by fraudsters, whereas this case is uncommon for real reviewers.  Furthermore, fraudster groups tend to have larger group size and higher support count (in that they share a large number of target products)---these features essentially reflect the size and density of the group's subgraph.

Incorporating time component, Beutel et al.~\cite{beutel2013copycatch} extended the group definition beyond graph connections by incorporating temporal closeness, and shows that group fraud (e.g., bots coordinating to post fake reviews) is temporally coherent as well and forms bipartite cores in a rearranged user-page bipartite network (Figure~\ref{fig:copycatch}). The existence of large temporally-coherent bipartite cores is highly suggestive of fraud.

\vspace{2mm}
Overall, opinion-based false information tends to be shorter, more exaggerated, and has more extreme ratings (1-stars and 5-stars). The fraudsters that create this false information give several ratings in a short time period (`bursty') and operate in a coordinated fashion (`lockstep').

\subsection{\textbf{Characteristics of fact-based false information}}

In this section, we discuss false information concerning facts with a single-valued truth.  This is different from information that may vary by someone's own opinion, e.g., their opinion about a particular product on Amazon.  Specifically, we discuss here the characteristics of hoaxes, rumors, and fake news.

\subsubsection{\textbf{Textual characteristics}}
There is a vast literature that studies fake news in social media. False information in the form of fake news is created in such a way to invoke interest and/or be believable to consumers. Various strategies may be used to deceive these consumers. Silverman~\cite{silverman2015lies} found that about 13\% of over 1600 news articles had incoherent headline and content body, for example, by using declarative headlines paired with bodies which are skeptical about the veracity of the information.

In a recent paper, Horne and Adali~\cite{horne2017fake} studied the textual characteristics of fake news using several sources of data: Buzzfeed fake news analysis articles~\cite{silverman2016analysis}, and articles from well known satire and fake news agencies (e.g., The Onion, Ending the Fed, and others). Reputed journalistic websites  were used for comparison.
The authors find interesting relations by comparing fake, satirical, and real news.
Below are two news article titles, one of which is fake. Can you identify the fake one?\footnote{First headline is fake.}

\begin{quote}
1. \textit{BREAKING BOMBSHELL: NYPD Blows Whistle on New Hillary Emails: Money Laundering, Sex Crimes with Children, Child Exploitation, Pay to Play, Perjury}
\end{quote}

\begin{quote}
2. \textit{Preexisting Conditions and Republican Plans to Replace Obamacare}
\end{quote}

Fake news tends to pack the main claim of the article into its title. The titles are longer but use fewer stopwords and more proper nouns and verb phrases, meaning that the creators tend to put as much information in the title as possible. The words used in the title are smaller and capitalized more often, to generate emphasis. Not surprisingly, titles of fake news and satire are very similar. In terms of the body content, fake news articles are short, repetitive, and less informative (fewer nouns and analytical words). They contain fewer technical words, more smaller words, and are generally easier to read. This allows the reader to skip reading the entire article, and instead just take information away from the title itself, which may be disparate from the rest of the content of the article.
Interestingly, Rubin et al.~\cite{rubin2016fake} studied satire news separately from the viewpoint of misleading readers into believing it is true, and also found that satirical articles pack a lot of information in single sentences. Thus, fake news articles are more similar to satirical ones than to real news---the bodies are less wordy and contain fewer nouns, technical and analytical words.
In addition, Perez et al.~\cite{perez2017automatic} also analyzed the textual properties of fake news using two datasets---one generated by Amazon Mechanical Turk workers and other one scraped on celebrity rumors from gossip websites. They found that fake news contains more social and positive words, is more certain, focuses more on present and future actions by using more verbs and time words.

But do people discuss false information differently from true information?
To answer this, Mitra et a.~\cite{mitra2017parsimonious} recently analyzed the language of how people discuss true and false information pieces using tweets of 1400 events. These events were part of their CREDBANK dataset, which used crowdsourcing to label ground truth credibility judgments~\cite{mitra2015credbank}.
Using LIWC categories~\cite{pennebaker2001linguistic}, they found that discussions around false information are marked with increased use of more confusion, disbelief, and hedging words which indicates skepticism among readers. Surprisingly, they found while more agreement words signaled high credibility, more positive sentiment words are associated with low credibility events. The latter is because it includes words like `ha', `grins', `joking' are positive sentiments, but instead mean mockery.
Their findings show that in addition to the text of the tweet itself, its surrounding discussion give important information to identify false information.

\begin{figure}
\centering

\includegraphics[width=0.3\textwidth]{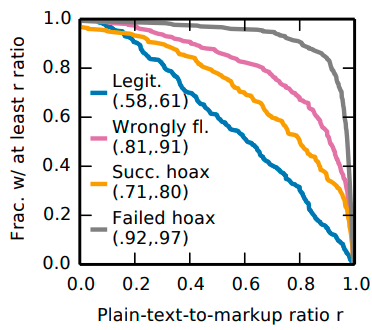}
\includegraphics[width=0.3\textwidth]{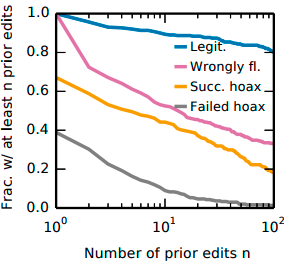}
\includegraphics[width=0.3\textwidth]{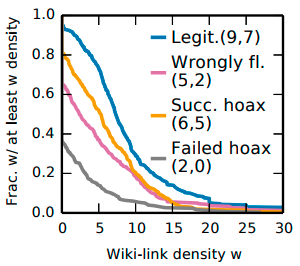}

\caption{Hoax articles (a) have lots of text, (b) fewer references, and (c) are created by newer accounts. Figure reprinted with permission from~\cite{kumar2016disinformation}.\label{fig:biased_ratings}}

\end{figure}

Hoaxes have similar textual properties as rumors. Kumar et al.~\cite{kumar2016disinformation} compared the content of hoax articles and non-hoax articles. They found that hoaxes were surprisingly longer compared to non-hoax articles (Figure~\ref{fig:biased_ratings}(a)), but they contained far fewer web and internal Wikipedia references (Figure~\ref{fig:biased_ratings}(b)). This indicated that hoaxsters tried to give more information to appear more genuine, though they did not have sufficient sources to substantiate their claims.

\subsubsection{\textbf{User characteristics}}
Several studies have shown that the characteristics of \textit{creators} of false information are different from those of true information creators. Kumar et al.~\cite{kumar2016disinformation} found that the creators of hoaxes have typically more recently registered accounts and less editing experience (Figure~\ref{fig:biased_ratings}(c)), suggesting the use of ``throw-away'' accounts. Surprisingly, non-hoax articles that are wrongly assumed to be hoaxes were also created by similar editors, meaning that they lack the skills to create well-written articles, which leads to others believing that the article is a hoax. 

In cases of rumors, Bessi et al.~\cite{bessi2016social} studied over 20.7 million tweets related to US presidential election, and identified users involved in tweeting as bots or honest users using a classification tool produced by Davis et al.~\cite{davis2016botornot}. Their analysis found that about one-fifth of content created and spread was by bots, showing that rumors are spread by automated accounts in short-bursts of time. Shao et al.~\cite{shao2017spread} came to similar conclusions in their experiments.

\subsubsection{\textbf{Network characteristics}}
Rumors and hoaxes can be related to other information in terms of what they say about others and what others say about it. Kumar et al.~\cite{kumar2016disinformation} quantified this for hoaxes on Wikipedia by measuring the connectedness of the different Wikipedia articles referenced in the hoax article. Intuitively, high connectedness indicates interrelated and coherent references.
The authors computed the clustering coefficient of the local hyperlink network of the article, i.e., the average clustering coefficient of the subnetwork induced by the articles referenced by the article. They found that hoax information has fewer references and significantly lower clustering coefficient compared to non-hoax articles. This suggests that references in hoaxes are added primarily to appear genuine, instead of adding them by need as legitimate writers do. 

Network characteristics of rumors are studied by analyzing the network of users that spread them and by creating co-occurrence networks out of false information tweets---these contain nodes of one or more types, such as URLs, domains, user accounts or hashtags, and use edges to represent the number of times they are mentioned in the same tweet together.
Using the user-user network, Subrahmanian et al.~\cite{subrahmanian2016darpa} found that some bot accounts that spread false information are close to each other and appear as groups in Twitter's follower-followee network, with significant overlap between their followers and followees.
Moreover, Bessi et al.~\cite{bessi2016social} conducted a k-core analysis of this follower-followee network and found that the fraction of bots increases steadily in higher cores, suggests that bots become increasingly central in the rebroadcasting network.
Using the co-occurrence network, Starbird~\cite{starbird2017examining} found that alternate media (false news) domains form tightly connected clusters, meaning that many users mention these domains together in their false information tweets.

\subsubsection{\textbf{Propagation characteristics}}
The spread of false information in social media makes it highly impactful. Several research studies have shown that only a small fraction of users are responsible for most of the spread, instead of being akin to a grass-roots movement. Gupta et al.~\cite{gupta2013faking} found that the top 30 users contributed towards 90\% of the retweets of fake images during hurricane Sandy on Twitter. Shao et al.~\cite{shao2016hoaxy} came to a similar conclusion in their study of about 1.3 million rumor tweets as well. Their analysis suggested that fake news spread was mostly dominated by a handful of very active users, whereas fact-checking of rumors was a more grass-roots activity with more conversation, and therefore slower.
This suggests that repetition and perseverance play an important role in the spread of false information. Since people tend to spread unverified claims~\cite{silverman2015lies,zubiaga2016analysing}, making false information believable may not be as important as persistently spreading it.

When false information spreads in social platforms, it spreads deeper compared to real news. In their study of rumor reshares on Facebook, Frigerri et al.~\cite{friggeri2014rumor} concluded that false information reshare cascades spread much deeper compared to that of true reference cascades. In other words, they are more likely to be reshared at greater depth and thus reach more people. One such reshare cascade is shown in Figure~\ref{fig:spread_fake_news}, with cascades colored by time. Additionally, Zeng et al.~\cite{zeng2016rumors} showed that information related to rumors, both supportive and denying, spread faster than non-rumors. Simulations conducted by Doerr et al.~\cite{doerr2012rumors} on realistic spread of simple rumors, on several graphs having the structure of existing large social networks, showed that even a rumor started by a random node on average reaches 45.6 million of the total of 51.2 million members within only eight rounds of communication. This is corroborated by the bursty behavior of rumors shown in several other research studies~\cite{silverman2016analysis,zubiaga2016analysing}.

\begin{figure}
\centering
\includegraphics[width=0.5\textwidth, height=2.5in]{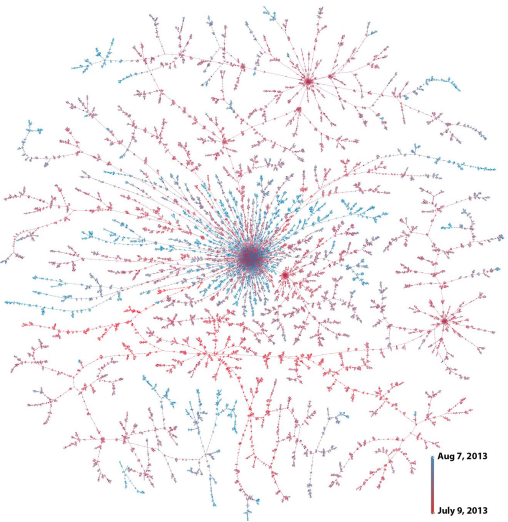}
\caption{Cascade of reshares of a Cabela's sporting goods store receipt attributing addition sales tax to ``Obamacare''. The coloring is from early (red) to late (blue). Reprinted with permission from~\cite{friggeri2014rumor}.\label{fig:spread_fake_news}}
\end{figure}

Several researchers have shown that false information spreads quickly, especially during its early stage.
Zubiaga et al.~\cite{zubiaga2016analysing} studied the entire lifecycle of true and false information as it spreads through social media, both before and after its veracity is checked. They collected 330 rumor threads with 4,842 tweets of nine popular cases, such as Charlie Hebdo shooting, and Michael Essien contracting Ebola.
Journalists then annotated the discussion threads of these rumors to quantify support expressed in tweets, i.e., their level of certainty (level of confidence indicated by the tweet) and  evidence (whether the tweet substantiates the rumor).
They found that the spread of false information occurs largely before it is even debunked. Tweets that supported unverified claims generated the most retweets, sparked by sudden bursts of retweets even during the first few minutes, with interest in the rumor decreasing substantially after its veracity is checked. During the initial spread of information, all users including normal users as well as reputed ones affiliated with news organizations, tend to tweet with a bias towards supporting unverified claims as opposed to denying them, irrespective of whether the information is later confirmed or denied.
Silverman~\cite{silverman2015lies} corroborates this finding. The level of certainty of tweets tends to remain the same before and after information is fact-checked, but users give more evidence before the rumor is fact-checked and less later on. These findings together further evidence the virality of popular false information during its initial phase.
Further, Vosoughi et al.~\cite{vosoughi2018spread} also showed that tweets about false information spread significantly farther, deeper, faster, and broader than those about true information.
This was observed for all categories of false information, such as politics, urban legend, science, business, and more.

While the above studies focus on spread of (false) information on a single platform, recent studies by Zannettou et al.~\cite{zannettou2017web} and Albright~\cite{albrightdata,albrightelection} mapped the false information ecosystem across social media platforms.
Zannettou et al.~\cite{zannettou2017web} studied the temporal relation between same information piece appearing on Twitter, Reddit, and 4chan platforms. They found that false information pieces are more likely to spread across platforms (18\% appear on multiple platforms) compared to true information (11\%). Moreover, false information appears faster across platforms than legitimate ones, and seems to `flow' from one to another, with Reddit to Twitter to 4chan being the most common route.
This spread across platforms is dangerous---Albright~\cite{albrightelection} studied the logs seven false information websites, and found that a whooping 60\% of incoming traffic was from Facebook and Twitter, and rest were from emails, search engines, messaging, or direct visits.
To study how these platforms connect to one another, Albright~\cite{albrightdata} crawled 117 false information websites and created a hyperlink network of domains that these websites refer to. He found that right-wing news websites link highly to other similar websites, thus supporting each other. Very surprisingly, YouTube was the most linked website overall, suggesting high use of multimedia content in conveying false information messages.

Thus, these studies have found that false information tends to propagate deeper and faster than true information, especially during the early stages of the false information. This happens on a single as well as across multiple platforms, and a handful of users are primarily responsible for this spread.

\subsubsection{\textbf{Debunking characteristics}}
Once false information spreads, attempts are made to debunk it and limit its spread. Recent research has shown that there is a significant time delay between the spread and its debunking. Zubiaga et al.~\cite{zubiaga2016analysing} found that true information tends to be resolved faster than false information, which tends to take about 14 hours to be debunked. Shao et al.~\cite{shao2016hoaxy} came to a similar conclusion---they found a delay of 10--20 hours between the start of a rumor and sharing of its fact-checking contents.

But once debunking information reaches the rumor spreaders, do they stop spreading it or does it `back-fire', as observed in in-lab settings~\cite{nyhan2010corrections} where corrections led to an \textit{increase} in misperception?
Several empirical studies on web-based false information suggest that debunking rumors is in fact effective, and
people start deleting and questioning the rumor when presented with corrective information.
Frigerri et al.~\cite{friggeri2014rumor} studied the spread of thousands of rumor reshare cascades on Facebook, and found that false information is more likely to be linked to debunking articles than true information. Moreover, once it is linked, it leads to a 4.4 times increase in deletion probability of false information than when it is not, and the probability is even higher if the link is made shortly after the post is created.
Moreover, Zubiaga et al.~\cite{zubiaga2016analysing} found that there are more tweets denying a rumor than supporting it after it is debunked, while prior to debunking, more tweets support the rumor.
Furthermore, Vosoughi et al.~\cite{vosoughi2018spread} showed that there is a striking difference between replies on tweet containing false information than those containing true information---while people express fear, disgust, and surprise in replies, true information generates anticipation, sadness, joy, and trust.
These differences can potentially be used to create early detection and debunking tools.

\vspace{2mm}
Overall, research on characterization of fact-based false information has shown that it tends to be longer, generates more disbelief and confusion during discussions, is created by newer and less experienced accounts that are tightly connected to each other, spreads deeper and faster in one and across multiple platforms, and gets deleted when debunking information spreads.

\begin{figure}[t]
\includegraphics[width=\textwidth]{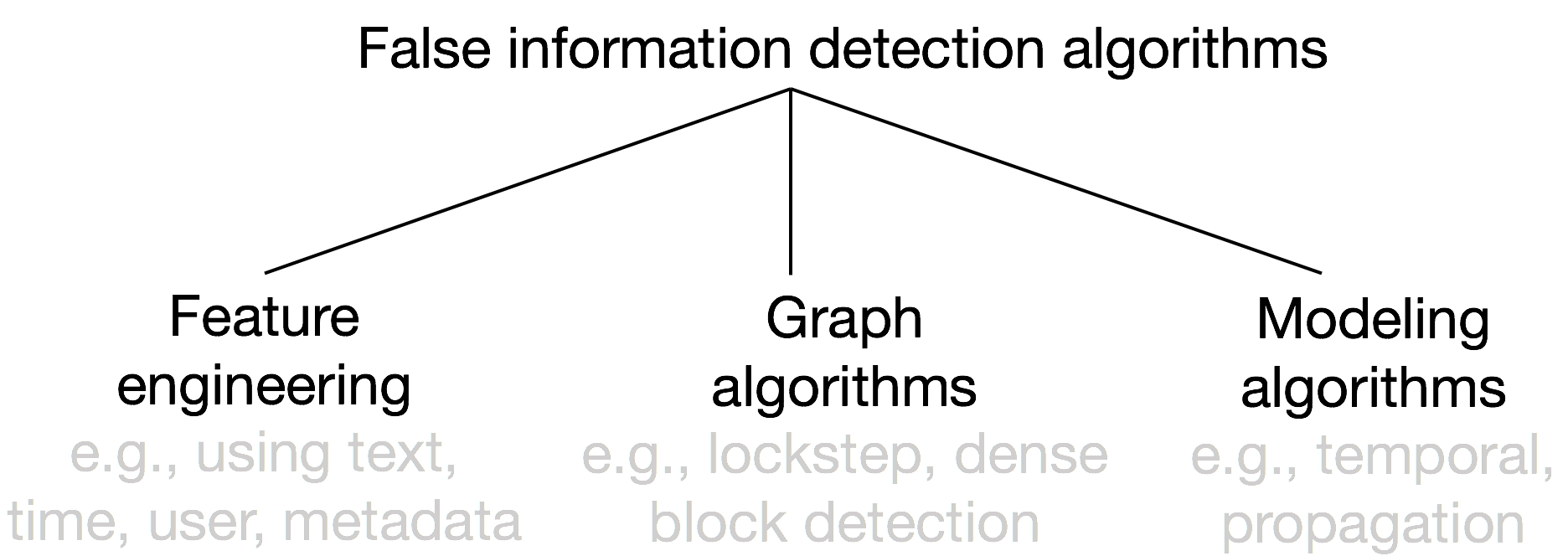}
\caption{Algorithms to detect false information, both opinion-based and text-based, can be broadly classified into (a) feature engineering, (b) graph algorithms, and (c) modeling algorithms.\label{fig:algorithm-category}}
\end{figure}

\section{\textbf{Detection of false information}}
\label{sec:detection}
In the previous section, we discussed a number of tell-tale signs and often-found characteristics of opinion-based and fact-based false information. In this section, we complement this information by discussing a number of approaches that researchers have employed to actually detect false information and those who spread it.

Algorithms to identify false information can be broadly categorized into three categories: feature engineering-based, graph-based, and modeling-based, as shown in Figure~\ref{fig:algorithm-category}.
The majority of algorithms are \textit{feature-based}, in that they rely on developing efficient features that individually or jointly are able to distinguish between true  and false information. These features are developed from the characterization analyzes that show the differences in properties of the two classes. These differences are then characterized by intelligently designed features. While we go into the details of some key research in feature-based detection, other papers that use features as described in Section~\ref{sec:characteristics} can directly be applied for detecting false information as well. Alternatively, \textit{graph-based} algorithms rely on identifying false information by targeting groups of users (spreaders) with unlikely high, lock-step coordination boosting a certain story (e.g., a botnet retweeting the same article in near-identical time). These algorithms try to identify dense blocks of activity in an underlying adjacency matrix. While these algorithms may be able to identify large-scale coordinated activity, small-scale or lone-wolf attacks are unlikely to be caught since the algorithms primarily focus on the largest dense blocks. Finally, \emph{modeling-based} algorithms work by creating information propagation models that emulate the empirical observation of edges and information spread. The intuition behind these algorithms is that since most information is true, it likely spreads a similar or unique way.  Thus, emulating this mode of information spread can pinpoint false information spread as anomalies which can then be verified and removed. In this section, we will look broadly at approaches belonging to these three classes for opinion-based and knowledge-based false information detection.
Table~\ref{tab:detect} categorizes the papers that develop detection algorithms into these three categories.

\begin{table}
\begin{tabular}{|c|c|c|}
\hline
\textbf{Algorithm category} & \textbf{Opinion-based false information} & \textbf{Fact-based false information}\\
&  \textbf{(fake reviews)}  &  \textbf{(false news and hoaxes)}  \\
\hline
\multirow{4}{*}{Feature-based} & 
Harris et al.~\cite{harris2012detecting},  Jindal et al.~\cite{jindal2008opinion},  &
Gupta et al.~\cite{gupta2013faking},  Horne et al.~\cite{horne2017fake},   \\

& Li et al.~\cite{li2011learning,li2014towards,li2015analyzing}, Lin et al.~\cite{lin2014towards},     & 
Kumar et al.~\cite{kumar2016disinformation}, Perez et al.~\cite{perez2017automatic},    \\
& Minnich et al.~\cite{minnich2015trueview}, Mukherjee et al.~\cite{mukherjee2012spotting, mukherjee2013yelp},    &  Qazvinian et al.~\cite{qazvinian2011rumor}, Rubin et al.~\cite{rubin2016fake},   \\
& Ott et al.~\cite{ott2011finding,ott2013negative}, Sandulescu et al.~\cite{sandulescu2015detecting}  & Wang et al.~\cite{wang2017liar},  Yang et al.~\cite{yang2012automatic}\\
& & \\

\multirow{3}{*}{Graph-based} & 
Akoglu et al.~\cite{akoglu2013opinion},  Hooi et al.~\cite{hooi2016birdnest},  &
Jin et al.~\cite{jin2016news},  Ruchansky et al.~\cite{ruchansky2017csi},  \\
&  Kumar et al.~\cite{kumar2018rev2},  Rayana et al.~\cite{rayana2015collective},  & Shu et al.~\cite{shu2017exploiting},  Tacchini et al.~\cite{tacchini2017some} \\
& Shah et al.~\cite{shah2016edgecentric},  Wang et al.~\cite{wang2011review} & \\
&& \\

\multirow{5}{*}{Modeling-based} & 
&  
Propagation: Acemoglu et al.~\cite{acemoglu2010spread},   \\
& Temporal: Xie et al.~\cite{xie2012review},  Ye et al.~\cite{ye2016temporal}  & Budak et al.~\cite{budak2011limiting}, Del et al.~\cite{del2016spreading},  \\
&Sentiment: Li et al.~\cite{li2014towards}  & Doerr et al.~\cite{doerr2012rumors},  Jin et al.~\cite{jin2013epidemiological},    \\
&& Nguyen et al.~\cite{nguyen2012containment},  Tripathy et al.~\cite{tripathy2010study}, \\
&& Wu et al.~\cite{wu2015false},  Zhao et al.~\cite{zhao2013sir}\\\hline

\end{tabular}
\caption{This table categorizes research research based on the type of false information detection algorithm. \label{tab:detect}}
\end{table}

The task of finding false information is one rife with challenges~\cite{subrahmanian2017predicting}. One of the major challenges arises from the imbalance in the population of two classes, false and true information In almost all cases, false information comprises only of a small fraction (less than 10\%) of the total number of instances. Moreover, false information is masqueraded to seem like truth, making it harder to identify. And finally, obtaining labels for false information is a challenging task. Traditionally, these are obtained manually by experts, trained volunteers, or Amazon Mechanical Turk workers.  The process requires considerable manual effort, and the evaluators are potentially unable to identify all misinformation that they come across. Any algorithm that is developed to identify false information must seek to address these challenges.

\subsection{\textbf{Detection of opinion-based false information}}

Here we look at the algorithms that have been developed in literature to identify opinion-based false information. Specifically, we look at the research on identifying fake reviews in online platforms using text, time, and graph algorithms.
Text-based algorithms primarily convert the textual information into a huge feature vector and feed that vector into supervised learning models to identify duplicate and fake reviews.
Graph-based algorithms leverage the user-review-product graph to propagate beliefs and to jointly model `trustworthiness' of users, reviews, and products.
Time-based algorithms employ time-series modeling and co-clustering along with feature engineering.
We will elaborate on these algorithms in the next few subsections.

\subsubsection{\textbf{Feature-based detection}}
As text is the primary source to convey (false) information on web platforms, it is one of the most widely studied component for fake review detection. Algorithms in this domain are based on \textit{feature engineering}, detecting \textit{duplicate reviews}, or a \textit{combination} of the two.
We primarily focus on text-based detection, as other features, such as user, graph, score, and time, are usually used in conjunction with text features in this task.

Several algorithms have been developed to efficiently identify duplicate reviews, with the notion that fraudsters give identical or near-identical reviews while genuine reviewers give more unique reviews.
Jindal et al.~\cite{jindal2008opinion} studied three major types of duplicates: different users reviewing the same product, same user reviewing different products, and different users on different products. They built a logistic regression model to detect fraudulent reviews incorporating rating and textual features such as review title and body length, sentiment, cosine similarity between review and product texts, and others, and achieved an AUC of 78\%. 
Similarly, Mukherjee et al.~\cite{mukherjee2012spotting} leveraged cosine similarity across a user's given reviews and across a product's received reviews in addition to rating and temporal features in an unsupervised generative Bayesian model to automatically discern separating features of truthful and fraudulent reviewers (AUC = 0.86).
Going beyond syntax, Sandulescu et al.~\cite{sandulescu2015detecting} studied the problem of detecting singleton review spammers by comparing both review syntax and semantic similarity in pairwise reviews per business, and marked reviews with high similarity as fraudulent. Syntactic similarity was measured using part-of-speech tags 
and semantic similarity using word-to-word distances in the WordNet synonyms database. This approach achieved F1-score between 0.5 and 0.7 on Yelp and Trustpilot customer reviews data, and suggests that intelligent fraudsters often duplicate semantically similar messages by replacing some words between their fake reviews with synonymous or similar words in order to avoid generating blatant duplicates and be caught.

However, more complex review fraud exists, where fraudsters put considerably more effort than just duplicating review text in order to write sophisticated, deceptive reviews. To get ground truth deceptive reviews, Amazon Mechanical Turk (AMT) workers are frequently employed.
The associated detection algorithms largely rely on text-processing and feature engineering for detecting such reviews.
Ott et al.~\cite{ott2011finding} collected 400 truthful and 400 positive-sentiment deceptive AMT-sourced reviews and trained Support Vector Machine (SVM) classifiers using a variety of feature sets, such as $n$-grams and LIWC features~\cite{pennebaker2001linguistic}. This achieved high 0.9 F1-score compared to human judges, who at best achieved a 0.7 F1-score.
In a followup work~\cite{ott2013negative}, negative sentiment deceptive reviews were studied, with additional 400 negative reviews from AMT.
Experiments showed that an SVM classifier trained on bigrams was again able to achieve strong performance in detecting such reviews, with a 0.86 F1-score.  The authors additionally studied classifier performance when training on positive sentiment reviews and testing on negative sentiment reviews, and vice versa---results showed that such heterogeneity between training and testing data produced considerably worse F1-score (roughly a 0.1--0.2 reduction) than the homogeneous case, indicating different statistical patterns in positive and negative sentiment deceptive reviews versus truthful reviews.

While AMT generated reviews are common to use, they lack domain expertise. To address that, Li et al.~\cite{li2014towards} collected additional deceptive reviews from domain experts such as employees at target venues.
The authors used $n$-gram features as in previous works and employ both Sparse Additive Generative Model (SAGE) and SVM classifiers to evaluate pairwise  and three-class classification (truthful customer vs deceptive Turker vs deceptive employee) performance (~65\% accuracy). Their results showed that distinguishing truthful customers from deceptive employees is somewhat more difficult than from deceptive Turkers.
Further, Li et al.~\cite{li2011learning} added sentiment, subjectivity, and pronoun usage features to the set. They then created a semi-supervised co-training algorithm that iteratively learns classifiers from review and reviewer features separately, and augments the training set in each successive iteration with the most agreed-upon and confidently scored reviews. This model achieves an F1-score of 0.63 on an Epinions review dataset.

A major aim of these systems is to aid humans in identifying fraud. Harris \cite{harris2012detecting} focused on the usefulness of these models to human judges. 
Equipping human judges with these simple summary statistics of reviews improved their manual classification accuracy by up to 20\% over the alternative (without), showing the effectiveness of augmented detection for humans at a cheaper computational cost.

\begin{figure}
\centering
\includegraphics[width=0.45\textwidth, height=2in]{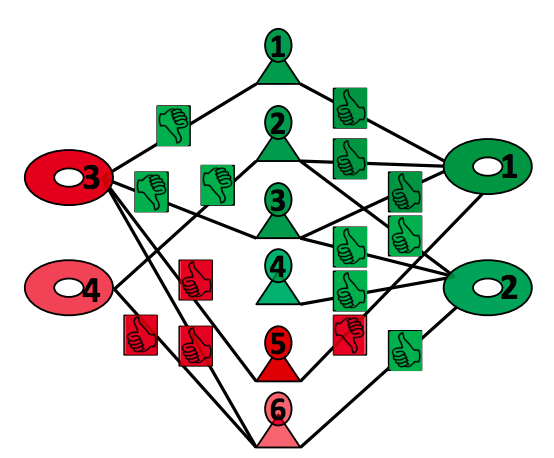}

\caption{Graph-based fake review detection algorithms are usually based on homophily, where good (green) users give positive ``thumbs up'' to other good products, while bad (red) users give negative ``thumbs down'' to them. The opposite is true for ratings given to bad products. Figure reprinted with permission from~\cite{akoglu2013opinion}.\label{fig:fraudeagle}}

\end{figure}

\subsubsection{\textbf{Graph-based detection}}
These algorithms leverage the rating graph for identifying fake edges. Nodes in the graph represent users and products, and edge from $u$ to $p$ represents a review by user $u$ to product $p$. Some algorithms also use features which may be available on nodes and/or edges.

Belief propagation on the rating graph is one common way to identify fraud.
Rayana et al.~\cite{rayana2015collective} used loopy belief propagation on the review graph network for identifying fake reviews, extending the idea of FraudEagle from Akoglu et al.~\cite{akoglu2013opinion}. The basic idea is presented in Figure~\ref{fig:fraudeagle}. These algorithms take a signed network, i.e. a network where the edges are converted to be positive (thumbs up) and negative (thumbs down), and employ the notion of homophily which suggests that most honest users give genuine positive ratings to good products and negative ratings to bad products, and vice-versa for bad fraudulent users. This is expressed as a Markov Random Field, where the joint probability $P(\textbf{y})$ of inferred labels $Y_i$ is a product of entity $i$'s prior beliefs $\phi_i(y_i)$ and its compatibility with labels of its neighbors $j$ represented as $\gamma_{ij}^s(y_i, y_j)$, with the compatibility matrix $s$. Mathematically,
$$ P(\textbf{y}) = \frac{1}{Z} \hspace{2mm} \Pi_{Y_i \in V} \phi_i(y_i) \hspace{2mm} \Pi_{(Y_i, Y_j, s) \in E^{\pm}} \gamma_{ij}^{s} (y_i, y_j) $$
This is solved using loopy belief propagation, with prior-based initialization and transfer of beliefs across the network till convergence. Based on this idea, Rayana et al.~\cite{rayana2015collective} combines belief propagation with feature values of nodes and edges as well. This SpEagle algorithm is highly accurate in identifying fake reviews (and users) in three Yelp fake review datasets, with area under the ROC curve scores around 0.78 on average.

Several algorithms have been developed for jointly modeling user, review, and product information, with applications to fake review detection. Wang et al.~\cite{wang2011review} uses the review network to measure trustiness of users $T(u)$, honesty of reviews $H(r)$, and reliability of stores $R(s)$, all of which lie between -1 and +1. For calculating $H(r)$ for a review $r$ given to product $p$, the trustiness of users $S_{r}^+$ and $S_{r}^-$ who gave similar and different scores, respectively, to product $p$ close in time to $r$. This is calculated as agreement score $A(r) = \sigma_{u_i \in S_{r}^+} T(u_i) - \sigma_{u_j \in S_r^-} T(u_j)$. Logistic functions are used to bound the scores in (-1, +1). The honest score $H(r)$ can then be used to identify fake reviews. The formulation is mutually interdependent as follows:
\begin{eqnarray}
T(u) & = & \frac{2}{1 + e^{-H(r)}} - 1\nonumber\\
H(r) & = & | R(s) | \hspace{2mm} (\frac{2}{1 + e^{-A(r)}} - 1) \nonumber \\
R(s) & = & \frac{2}{1+e^{-\Sigma_{(u,s) \in In(s)}T(u)}} - 1\nonumber
\end{eqnarray}
The authors tested the algorithm to identify fake reviewers, which is a closely related problem, and get a precision of 49\% on the 100 users with the smallest trustiness scores.

Closely related to the previous algorithm is Rev2 by Kumar et al.~\cite{kumar2018rev2}, which is also an iterative algorithm which calculates reviewer fairness, rating reliability and product goodness scores. The algorithm is based on the intuition that fraudulent review writers are typically unfair, in that they give unreliable rating scores to products that differ largely from the product's average score. Reliable reviewers give ratings that are close to the scores of other reliable reviewers.
This algorithm incorporates user and product features by merging scores from a prior algorithm called Birdnest~\cite{hooi2016birdnest}, and uses Bayesian priors for addressing cold start problems, i.e., judging the quality of users and products that only have a few ratings. This formulation is also interdependent, and the rating reliability is used to identify fake reviews. This algorithm achieves an AUC of over 0.85 for identifying fraudulent reviewers.

Several graph based algorithms have been developed to identify fraudulent nodes in review networks, using edge distributions~\cite{shah2016edgecentric,hooi2016birdnest}, dense block detection~\cite{beutel2013copycatch,jiang2014inferring}, co-clustering~\cite{beutel2014cobafi}, and more. This problem is closely related to identifying fake reviews, as the intuition is that by identifying fraudulent users, one can identify remove all their reviews and eliminate fake reviews. However, while these techniques may work for identifying bad users, these may not work well as-is in fake review detection because of two reasons: first, not all reviews by fraudulent users are necessarily fake~\cite{kumar2018rev2} (for example, the user might aim to camouflage themselves by giving a few genuine reviews) and second, not all fake reviews are given by fraudulent users, which would hinder recall for fake review detection. Thus we do not focus on these algorithms in detail.

\begin{figure}
\centering
\includegraphics[width=0.45\textwidth]{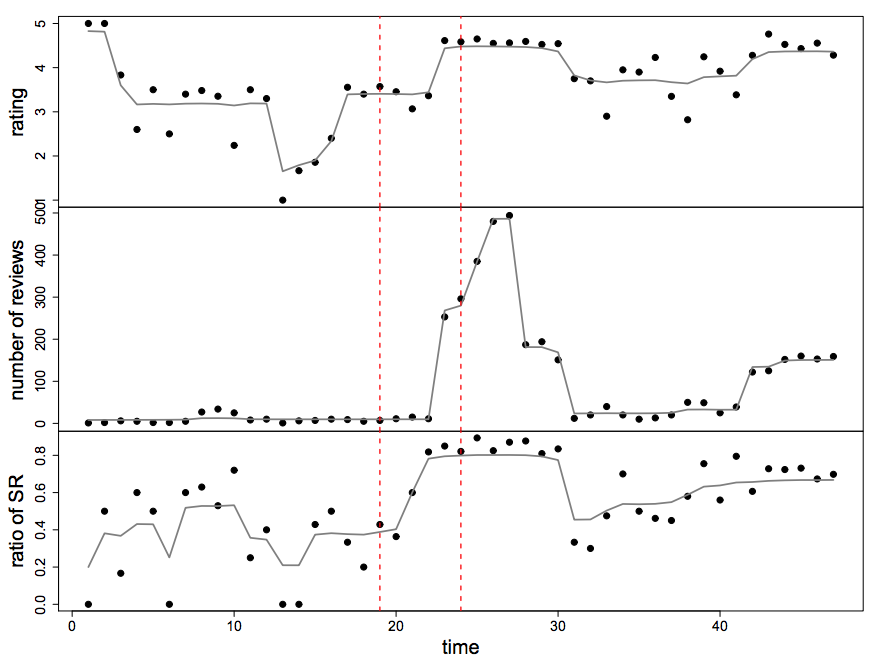}
\caption{Figure showing suspicious rating time range (between red line) by showing bursty increase in rating (top), number of reviews (middle) and ratio of singleton review (bottom) in a coordinated manner. Figure reprinted with permission from~\cite{xie2012review}.   \label{fig:xie2012review}}
\end{figure}

\subsubsection{\textbf{Detection with temporal modeling}}
This category includes algorithms that primarily using rating time information for identifying fake reviews. These approaches include a combination of feature engineering and modeling based on time series analysis, correlation, and co-clustering.

Ye et al.~\cite{ye2016temporal} considered the review data of each product as a stream of reviews, bucketed into temporal windows. For each temporal window, the authors consider a number of time series reflecting different properties: inter-rating time entropy, rating score entropy, average rating, review count, number of positive and negative reviews, and more. 
They created a LocalAR algorithm to identify anomalies in the time-series signal, which involves treating a single signal as the ``lead'' and using the rest of the signals as ``supporting'' ones. An abnormality score at a timestep in the lead signal is defined as the residual between an empirically observed value and the forecasted value based on a local autoregressive (AR) model from previous timesteps. AR models generally take the following form:

\begin{displaymath}
X_t = \sum_{i=1}^{k} c_i X_{t-i} + \epsilon
\end{displaymath}

The authors keep track of the distribution $D(S|T,P)$ of abnormality scores $S$ over all timesteps $T$ and products $P$, and define a threshold to flag abnormal timesteps, estimated as the percentage of expected anomalies using Cantelli's inequality.
Upon finding anomalous points in the lead signal by this threshold, the algorithm turns to supporting signals for corroboration.
For timesteps near only those flagged in the lead signal, the approach computes a local AR model on the supporting signals and calculates residual squared-error between estimates and empirically observed values at surrounding timesteps.
If the residual error is above the abnormality threshold, the timestep is flagged as suspicious. These flags are integrated across the multiple signals  using summary statistics like proportion of anomalies at a timestep. 
This LocalAR algorithm shows success on two case studies of bursty opinion spam on Flipkart data.

Along similar lines, Xie et al.~\cite{xie2012review} created CAPT-MDTS (Correlated Abnormal Patterns Detection in Multidimensional Time Series) based on burst detection in time-series. Specifically, the authors aim to find time periods during which a product is ``under attack'' by review fraudsters.
The proposed approach involves detecting periods of time where there are correlated bursts in multiple time series reflecting average rating, review count, and proportion of singleton reviews. An illustration is given in Figure~\ref{fig:xie2012review}. The burst detection algorithm is built upon a variant of the longest common subsequence (LCS) problem which allows for two or more sequences (in this case, time series) to be considered ``common'' if they are approximately the same.

Furthermore, Li et al.~\cite{li2015analyzing} focused on detecting review fraud using both rating and spatiotemporal features in a supervised setting.  The authors show that high average absolute rating deviation and high ``average travel speed'' (spatial distance between two subsequently reviewed venues divided by time between reviews) are suspicious, in addition to high distance between the registered location of the reviewer's account and the venue he/she is reviewing.

\vspace{2mm}
Overall, algorithms developed to identify opinion-based false information rely primarily on the information text, the entire user-review-product graph, and temporal sequence of reviews to identify individual and group of false reviews. Additional information, such as user properties, help as well. These algorithms have been developed and tested in a wide variety of platforms and datasets, and are efficient (high precision and AUC scores) in identifying fake reviews.

\subsection{\textbf{Detection of fact-based false information}}

In this part, we will look at the algorithms to detect hoaxes, fake news, and rumors in social media.
These algorithms can be categorized into two major categories: feature engineering based and propagation based.
Similar to opinion-based feature engineering methods, here features are created from their textual properties, their relation to other existing information, the properties of the users interacting with this information (e.g., the creator), and propagation dependent features (e.g., number of users that reshare a tweet).
Feature-based algorithms have been used to identify various different types of malicious users and activities, such as identifying bots~\cite{subrahmanian2016darpa}, trolls~\cite{cheng2015antisocial}, vandals~\cite{kumar2015vews}, sockpuppets~\cite{kumar2017army}, and many more.

Fact-based false information propagates through social networks, as opposed to opinion-based false information. Thus, propagation based algorithms model how true information propagates in these networks and anomalies of these models are predicted as false information. Some algorithms also create separate models for true and false information propagation. Alternatively, propagation structures and information can be fed into machine learning models for prediction as well.

We will first discuss feature based algorithms (Section~\ref{sec:feature-text}) and then explain propagation based models in Section~\ref{sec:propagation}.

\subsubsection{\textbf{Feature-based detection}}

\noindent\textbf{Text-based features:}\\
\label{sec:feature-text}
Text-analysis is core to identifying misinformation as the information being conveyed is primarily textual content. Similarly to opinion-based textual detection methods, research papers in this category are predominantly feature-based, where features can broadly be categorized as either stylometric (e.g., number of characters in a word), psycholinguistic (e.g., LIWC~\cite{pennebaker2001linguistic}), or complexity-oriented (e.g., readability indices).

One of the first studies on identifying rumors on Twitter was done by Qazninian et al.~\cite{qazvinian2011rumor}. They collected manual annotations for over 10,000 tweets, and developed three categories of features to identify the false tweets---primarily based on content (unigram, bigrams, and part-of-speech), but also used user information (whether user has previously posted false information), and Twitter-specific information (hashtags and URLs).
These features were converted into their log-likelihood ratio of being from the true or false class based on their distribution in the training data, and the combined score was used for classification. This model achieved a mean average precision score of 95\%, indicating near-perfect classification. Individually, content features performed the best, followed by network features, and lastly hashtag and URL based Twitter features. Content based features were also the best performing features in Gupta et al.~\cite{gupta2013faking}, which focused on fake tweet detection as well.

A recent study on false news detection by Perez-Rosas et al.~\cite{perez2017automatic} shows the changing effectiveness of text-based features. They collected a large dataset of false information generated by Amazon Mechanical Turk (AMT) workers and another dataset of celebrity fake news from gossip websites. They used a huge set of text based features for classification, consisting of $n$-grams, punctuations, LIWC, readability, and syntax features.
Since the experiments were conducted on a balanced dataset, the baseline accuracy is 50\%, and the combined set of features achieves an average accuracy of 74\%. Unsurprisingly, cross-domain analysis, i.e., training model on one dataset and testing on another, dropped the accuracy to 61\%.
Within the same dataset, but training and testing on different domains of fake news (e.g., technology, education, etc.) the performance lies between 75\% and 85\%, indicating that knowledge can be transferred from one domain to another with reasonable performance.

Similarly, in their study to identify fake from real news from text, Horne and Adali~\cite{horne2017fake} achieved accuracies between 71\%--78\%. Identifying fake news from satire was more difficult than identifying fake news from real news, indicating that the former two are written quite similarly.
In their experiments, they observe that the news body text is less informative than the news title text in distinguishing real from fake information. This is an important finding and opens avenues of future research to quantify the dissonance between the title and body of an information piece, and using it as an indicator of false information. 

The above studies show an alarming trend. In the earlier research papers, content-based detection performed well, but more recent research has shown that content alone is not good enough. This suggests a trend that malicious users and content creators are adapting and becoming more aware of how to create more genuine-looking false information which fools automated classifiers.

\begin{figure}
\centering

\includegraphics[width=0.4\textwidth, height=1.8in]{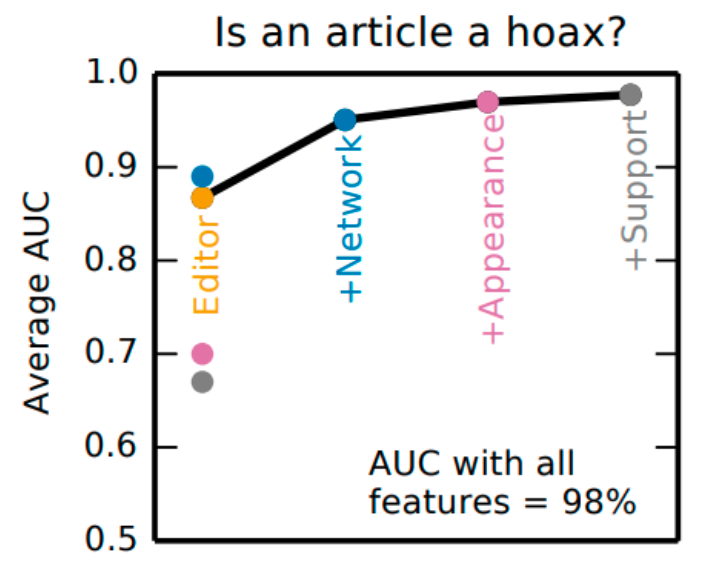}

\caption{Plot showing accuracy of hoax detection using different feature sets. Reprinted with permission from ~\cite{kumar2016disinformation}.\label{fig:hoax_detection}}

\end{figure}

\vspace{2mm}
\noindent \textbf{User, network, and metadata based detection:}\\
These detection models derive features from several different aspects of data which we describe below. We start by looking at the features developed to identify hoaxes in Wikipedia. Kumar et al.~\cite{kumar2016disinformation} developed four different categories of features to identify hoaxes: (i) \textit{appearance features}, that measures the length of the article, number of references, ratio of text to Wikipedia markup and so on, (ii) \textit{network features}, that measures the relation between the references of the article in the Wikipedia hyperlink network, (iii) \textit{support features}, quantifying the number of previous occurrences of the article title in other articles, and the time since its first occurrence to the time the article is created, and (iv) \textit{creator features}, i.e., properties of article creator in terms of its previous edit count, time since registration, and so on.
Figure~\ref{fig:hoax_detection} shows the performance of the features using a random forest classifier. Individually, creator and network features perform equally well (AUC = 0.90)
Further, the performance can be improved significantly when other sets of features are incrementally added. The combination of all four categories of features gives an AUC of 0.98, indicating a near perfect classifier. This shows that in order to identify false information, one needs to look beyond its appearance and dig deeper into who created the piece of false information and how it relates to existing information---simply looking at its appearance is not as effective.

In the domain of fake Twitter images during disasters, Gupta et al.~\cite{gupta2013faking} used user features (number of friends, account age, etc.), tweet features (number of words, sentiment, POS, LIWC, etc.), and metadata features (number of hashtags, mentions, URLs, retweets).
The dataset had 11,534 tweets, half of which were fake and other half were real, and decision trees were used for classification. User features alone had close to random accuracy of about 53\%, while tweet features alone got near perfect accuracy of 97.7\%, indicating that in the propagation of false information, what the tweet says is more important than the user who tweets it.

Thus, feature engineering has proven successful in identifying fake from true rumors and hoaxes, primarily using features derived from the text, user, network, and other metadata.
These algorithms typically have performance numbers in high 80s and 90s, showing that they are effective and practically useful.

\begin{figure}
\centering
\includegraphics[width=0.25\textwidth, height=1.2in]{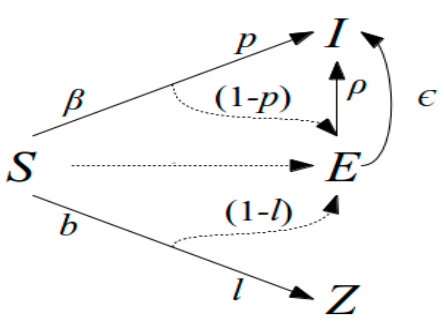}
\includegraphics[width=0.3\textwidth, height=1.2in]{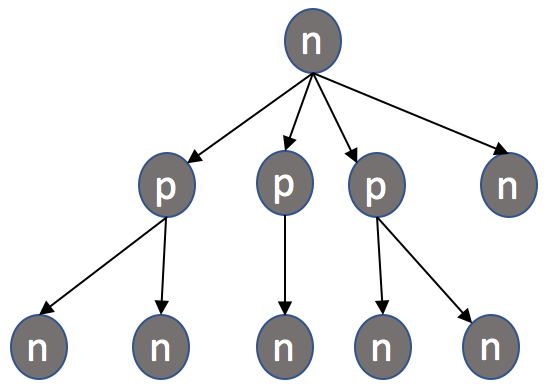}
\includegraphics[width=0.3\textwidth, height=1.2in]{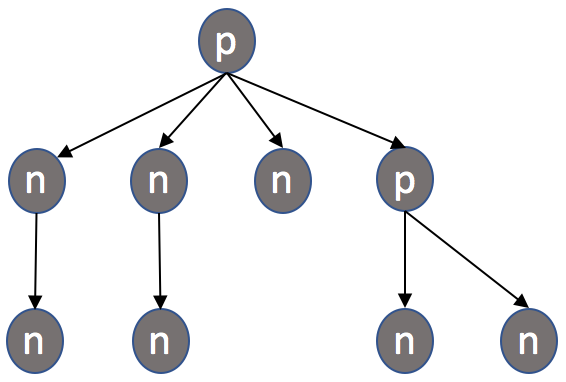}
\caption{(a) SEIZ model of information propagation. Reprinted with permission from ~\cite{jin2013epidemiological}. (b--c) Common information propagation structure or motifs for (b) false information, and (c) real information. Adapted from~\cite{wu2015false}. \label{fig:message-prop} }

\end{figure}

\subsubsection{\textbf{Detection using propagation modeling}}
\label{sec:propagation}
The spread of information across social media adds another dimension to use for identification of true information from false information.
A common way of using propagation information is to create models that broadly serve two purposes: to closely model the spread of (false) information, and to find ways to prevent its spread.

\noindent\textbf{Creating true information propagation models:}\\
Acemoglu et al.~\cite{acemoglu2010spread} presented one of the first models to simulate propagation of information in social media. They considered information spread as exchange of belief about information (e.g., supporting a political candidate), and theoretically find the cases in which false information survives in the network. Nodes in the network are considered to be either normal or forceful. When two normal nodes interact, each of them adopts an average of their existing beliefs. But interactions with forceful nodes only change the belief of the normal node, while forceful node only slightly updates its beliefs. With this interaction model, simulations showed that belief about false information dies out when the social network is well connected and normal nodes interact with each other a lot. On the other hand, \textit{echo chambers} are formed and false information prevails when there are several forceful nodes who update their own belief from the beliefs of nodes they previously influenced. This model suggests increasing the number of normal nodes and increasing their connectivity with each other may be a way to reduce false information propagation.

More recently, Jin et al.~\cite{jin2013epidemiological} created a false information propagation model which they called SEIZ. Similar to the standard SIS model, each node in SEIZ model lies in one of four states---susceptible (S), exposed (E), infected (I), and skeptical (Z), as shown in Figure~\ref{fig:message-prop}(a). Initially, nodes are susceptible (state S). When they receive information, they can transition to states I or Z with probabilities $\beta$ and $b$, respectively. These transitions may only be successful with some probabilities $p$ and $b$, respectively, otherwise the nodes transition to state E.
The following four equations explain the transitions according to this SEIZ model:
\begin{eqnarray}
\vspace{-2mm}
\frac{d[S]}{dt} & = &-\beta S\frac{I}{N} -bS\frac{Z}{N} \nonumber\\
\frac{d[E]}{dt} & = & (1-p)\beta S\frac{I}{N} + (1-l)bS\frac{Z}{N} - \rho E \frac{I}{N} - \epsilon E \nonumber\\
\frac{d[I]}{dt} & = & p \beta S \frac{I}{N} + \rho E \frac{I}{N}  + \epsilon E  \nonumber\\
\frac{d[Z]}{dt} & = &   lbS\frac{Z}{N} \nonumber
\end{eqnarray}
The model parameters are learned by training on real true and false information propagation data, and these are used for prediction of false information.
A metric measuring the rate of users entering state E to leaving it is predictive of false information---a high value indicates true information while a low score indicates a rumor, as shown by their case study of eight rumors.

\noindent\textbf{Incorporating propagation information in machine learning models:}\\
Apart from model creation, propagation information and propagation structure can both augment existing machine learning frameworks. For example, Yang et al.~\cite{yang2012automatic} added propagation features such as number of replies and number of retweets to the standard set of features used for their classification tasks, such as content, client (device type), user, and location, to identify false information spread on the Sina Weibo network. Using this feature set with an SVM classifier gave an average accuracy of 78\%.

A more structured way of using propagation information was created by Wu et al.~\cite{wu2015false}, who also studied 11,466 rumors on Sina Weibo by using propagation information.
Each thread of information was represented as a tree, with the original message as the root and its replies as children, and so on. This is shown in Figure~\ref{fig:message-prop} (b) and (c) for false and true information, respectively.
Popular nodes, i.e., ones with at least 1,000 followers, are denoted as $p$ and others as $n$, to understand if popular nodes boost false information.
The authors observe that false information is usually started by a normal user, then reported and supported by some opinion leaders, and then finally reshared by a large number of normal users (Figure~\ref{fig:message-prop}(b)). On the other hand, true information is posted by opinion leaders and then reposted directly by normal users (Figure~\ref{fig:message-prop}(c)).
With features representing the propagation structure, user features, average sentiment, doubt, surprise and emotion, an SVM classifier achieved 91\% accuracy. Further, early detection of false information achieved 72\% accuracy even without any propagation information, and 88\% with propagation information of its first 24 hours.

\noindent\textbf{Mitigation by modeling true and false information:}\\
Previously we described propagation models spreading false information. Here, we consider models that model the spread of both true and false information simultaneously, where success is measured as the number of users saved from accepting false information. The aim of these models is to create mitigation strategies to reduce and prevent the spread of false information. Several such models have been developed.
Tripathy et al.~\cite{tripathy2010study} created two models in which true information (anti-rumor) is spread after false information(rumor) starts to spread, based on real data. In one model, truth is spread \textit{n} days after falsehood to simulate real-world observed time-lag, and in the second model, truth is spread by some special nodes (e.g., official accounts) as soon as they receive false information. They conducted experiments with Twitter and simulated networks, and find that there is a super-linear relationship between the lifetime of a rumor and delay of its detection. When the special nodes detect and spread anti-rumors, it reduces the average lifetime of rumors by over 60\%.

Similarly, Budak et al.~\cite{budak2011limiting} presented an independent cascade model called Multi-Campaign Independence Cascade Model (MCICM). The model contains a rumor campaign and a true information `limiting' campaign spreading through the network. Each node, whenever infected with true or false information, spreads its belief to its neighbors, which accept the information with some probability. Their algorithm learns the model parameters, even with missing data. Their simulations show that a 30\% increase in delay in starting the true information spread reduces its reach by 80\% of the population.
More recently, Nguyen et al.~\cite{nguyen2012containment} created another model with both linear threshold and independent cascade, and found that when true information can only be spread by a few nodes, it is most effective to do it via highly influential nodes in large communities. When more nodes can be selected, influential nodes in smaller communities are more effective in preventing the spread of false information. This method is 16--41\% better than other methods. Related models have also been developed~\cite{zhao2013sir}.

Thus, several propagation models have been developed that capture the spread of true and false information on social media. These models are used independently or in conjunction with other machine learning algorithms to identify false information. These algorithms are effective in detecting spread of rumors, and their simulations suggest rumor mitigation strategies.

\vspace{2mm}
Overall, several categories of false information detection algorithms have been developed for opinion-based and fact-based false information.
The common category of algorithm is feature-based, which converts observations into feature vectors, derived from text, user, network, and metadata.
Graph-based algorithms have primarily been developed for opinion-based false information (e.g., fake reviews), and identify dense block of users or information, potentially also occurring in bursty short time period.
Temporal modeling algorithms use time-series analysis, and co-clustering on one or more of time evolving properties of information (e.g., number of ratings per day) to identify opinion-based false information as well.
Fact-based information that spreads (e.g., rumors) is also detected by creating true and false information propagation models.
All these types of algorithms perform well in their respective datasets to identify false information, and usually achieve a high accuracy, precision, or AUC score in the 80s or 90s.

\section{\textbf{Discussions and Open Challenges}}
In this survey, we took a comprehensive view on mechanisms, rationale, impact, characteristics, and detection of three types of false information: fake reviews, hoaxes, and fake news.

Several algorithms have been developed for detection in different domains. However, they are not directly comparable to each other due to the lack of large-scale publicly available datasets of false information, spanning fake reviews, hoaxes, and social media rumors. This prevents a benchmark comparison between different categories of algorithms. Such datasets are needed to understand the advantages and disadvantages of different algorithms, and collectively improve the state of the art. Some recent datasets, such as from Buzzfeed~\cite{silverman2016analysis}, LIAR~\cite{wang2017liar}, and CREDBANK~\cite{mitra2015credbank}, and FakeNewsNet~\cite{shu2017fake,shu2017exploiting}, have been created but standardized comparison of existing algorithms on these datasets has not been conducted.

The next generation of false information will be fueled by the advancements in machine learning. Recent research has shown that machine learning models can be built to create genuine looking text~\cite{yao2017automated}, audio~\cite{mukhopadhyay2015all}, images and videos~\cite{suwajanakorn2017synthesizing}. With further development of such techniques, it will become increasingly difficult for readers to identify false from true information. Techniques to separate the two using standard signals, like text, user information, group-level interaction, time, and more will need newer reforms for combating this new wave.

There are several open avenues of research, spanning areas of machine learning, natural language processing, signal processing, information retrieval, big data analysis, and computer vision for studying, characterizing, detecting and preventing false information on the web and social media:

\noindent\textbf{Semantic dissonance detection:}
Some smartly created false information pieces cite references to look credible, but the reference may not reflect what the information piece says. Often, the summary of some information (e.g. the headline of a news information piece) may not convey the same message as the main information itself. These tactics leverage human laziness and aversion to fact-checking. Natural language processing algorithms for semantic dissonance estimation can help identify such deceptive sources. 

\noindent\textbf{Fact-checking from knowledge bases:}
Fact-based false information can be checked by matching it against a knowledge base of complete information. This direction poses numerous challenges. The first is successfully creating and maintaining this knowledge-base, and ascertaining data quality. Next, natural language understanding and information extraction techniques must be developed to automatically extract information from free-form natural text. Finally, we would require capable information matching algorithms in order to check if the extracted information matched with the existing information in the knowledge base.

\noindent\textbf{Fact-checking using crowdsourcing:}
Readers express different emotions, such as skepticism, when interacting with false information as compared to true information~\cite{vosoughi2018spread}. They may also ``report'' such posts. Manual fact-checking of all stories is not feasible and such stories can be created to bypass existing detection filters, which is where crowdsourced signals can be useful. These algorithms can help in early detection of fake news~\cite{kim2018leveraging} and resource allocation of fact-checkers.

\noindent\textbf{Multimedia false information detection:}
As demonstrated by recent research, fabricated and manipulated audio~\cite{mukhopadhyay2015all}, images and videos~\cite{suwajanakorn2017synthesizing} can be developed using learning technologies. Research topics in these directions include developing signal processing, computer vision, and data analysis techniques to identify signature characteristics of fabricated or manipulated multimedia, and developing machine learning algorithms for their detection.

\noindent\textbf{Bridging echo chambers:}
The formation of social media echo chambers fuels the presence and spread of false information. One strategy to combat false information is to bridge conflicting echo chambers, so that opposing information can be exchanged and considered. Data-driven models of effective means of bridging these echo chambers/filter bubbles are needed. At the same time, further research is required in order to effectively present opposing beliefs to readers in order to reduce polarization.

\noindent\textbf{Adversarial creation of false information:}
Malicious users that try to create and spread false information are actively involved in the process. They can adapt their future behavior based on the counter-measures that are being taken to detect and prevent their current behaviors. Therefore, research in the direction of dealing with adaptive adversaries is promising in mitigating the impact of false information.

\noindent\textbf{Mitigation of false information:}
Reducing the damage of false information is an essential direction that is open for research. Very recent research has shown that educating people against possible manipulation strategies used in false information is effective in improving human detection skills~\cite{van2017inoculating}. Further research in finding effective educational strategies to ``vaccinate'' people against believing false information, and how to scale these strategies to millions of users that use social platforms is necessary.

\section*{Acknowledgement}
The authors would like to thank Prof. Jure Leskovec for comments on an early draft of this survey, and Prof. Jiliang Tang and Dr. Charu Aggarwal for inviting us to write an earlier version of this survey for their book. We thank the authors for giving us permission to reprint their figures.

\bibliographystyle{plain}
\bibliography{bib/neil,bib/srijan}
\end{document}